\documentclass{article}
\usepackage{kr}

\usepackage{times}
\usepackage{soul}
\usepackage{url}
\usepackage[hidelinks]{hyperref}
\usepackage[utf8]{inputenc}
\usepackage{multirow}
\usepackage{tabularx}
\usepackage[small]{caption}
\usepackage{subcaption}
\usepackage{graphicx}
\usepackage{amsmath}
\usepackage{amsthm}
\usepackage{amsfonts} 
\usepackage{amssymb}
\usepackage{booktabs}
\usepackage{algorithm}
\usepackage{algorithmic}
\usepackage{textcomp}
\usepackage{subcaption}
\usepackage{array}
\usepackage{mathtools}
\usepackage{xspace}
\usepackage{multirow}
\usepackage{subfiles}
\usepackage{tikz}
\usetikzlibrary{trees,automata,positioning}
\usepackage{enumitem}

\newcommand{\rj}[1]{\textcolor{red}{#1}}

\usepackage{amsmath,amsfonts,amssymb}
\usepackage{bm}
\usepackage{tikz}
\def\mergeset{\textsc{eq}}
\def\suppset{\textsc{supp}}
\def\vioset{\textsc{abs}}
\def\viorset{\textsc{viol}}

\def\activepairs{\mathbf{actP}}

\newcommand{\StrPos}{\mathsf{Cells}}

\newcommand{\wrt}{w.r.t.~}

\newcommand{\ie}{i.e.~}

\newcommand{\tid}{\ensuremath{\mathsf{tid}}}
\newcommand{\ours}{$\textsc{ASPen}^{\!+}$\,}
\newcommand{\prev}{\textsc{ASPen}\,}

\newcommand{\LACE}{\textsc{Lace}\,}
\newcommand{\laceEXT}{\textsc{Lace}$^+$\xspace}

\newcommand{\tbl}[1]{Table~\ref{#1}}
\newcommand{\fig}[1]{Figure~\ref{#1}}
\newcommand{\sect}[1]{Section~\ref{#1}}

\def\dcons{\mathsf{Dom}}

\def\TID{\mathbf{TID}}

\newcommand{\Author}{\text{Author}}
\newcommand{\Awrd}{\text{Awarded}}
\newcommand{\aid}{\ensuremath{\mathsf{aid}}}
\newcommand{\name}{\ensuremath{\mathsf{name}}}
\newcommand{\dob}{\ensuremath{\mathsf{dob}}}
\newcommand{\pob}{\ensuremath{\mathsf{pob}}}
\newcommand{\awrd}{\ensuremath{\mathsf{awrd}}}

\newcommand{\aspauthor}{\mathtt{author}}
\newcommand{\aspawarded}{\mathtt{awarded}}

\newcommand{\MaxSOL}{\mathsf{OptSol}}
\newcommand{\SOL}{\mathsf{Sol}}

\newcommand{\SRV}{\Gamma_s^v}
\newcommand{\HRV}{\Gamma_h^v}

\newcommand{\Obj}{\bm{\mathsf{O}}}
\newcommand{\ObjD}{\mathsf{Obj}}
\newcommand{\Val}{\bm{\mathsf{V}}}
\newcommand{\tidD}{\bm{\mathsf{TID}}}

\newcommand{\linkO}{\mathsf{EqO}}
\newcommand{\linkV}{\mathsf{EqV}}

\newcommand{\dom}[1]{\mathsf{dom}(#1)}

\def\harrow{\Rightarrow}
\def\sarrow{\dashrightarrow}
\newcommand{\tup}[1]{\langle #1\rangle}

\newcommand{\E}{\Sigma}

\newcommand{\D}{\mathcal{D}}

\renewcommand{\S}{\mathcal{S}}

\newcommand{\magellan}{Magellan\,}
\newcommand{\jedai}{JedAI\,}

\newcommand{\imdb}{\textit{Imdb}\,}
\newcommand{\corimdb}{\textit{ImdbC}\,}
\newcommand{\music}{\textit{Mu}\,}
\newcommand{\cormusic}{\textit{MuC}\,}
\newcommand{\cellmusic}{\textit{MuCC}\,}
\newcommand{\poke}{\textit{Poke}\,}

\newcommand{\tground}{\mathbf{t_g}}

\newcommand{\tsolve}{\mathbf{t_s}}

\newcommand{\ter}{\mathbf{t}_\text{o}}
\newcommand{\enumc}{\mathbf{\#e}}
\def\tenumone{\mathbf{t_{s}^{1}}}
\def\tenumn{\mathbf{t_{s}^{n}}}

\def\fone{\mathbf{F_1}}
\def\avgfone{\mathbf{\overline{F}_1}}
\def\avgp{\mathbf{\overline{P}}}
\def\avgr{\mathbf{\overline{R}}}

\newcommand{\ASet}[1]{\textit{SM}(#1)}
\newcommand{\aspenc}{\ensuremath{\Pi_{(D, \Sigma)}}\xspace}

\newcommand{\defneg}{\mathtt{not}}

\newcommand{\simpred}{\mathtt{sim}\xspace}

\newcommand{\emptypred}{\mathtt{empty}}

\newcommand{\aspeqo}{\mathtt{eqo}\xspace}
\newcommand{\aspeqv}{\mathtt{eqv}\xspace}
\newcommand{\aspneqo}{\mathtt{neqo}\xspace}
\newcommand{\aspneqv}{\mathtt{neqv}\xspace}

\newcommand{\enumi}{\textbf{i)}}
\newcommand{\enumii}{\textbf{ii)}}
\newcommand{\enumiii}{\textbf{iii)}}

\newcommand{\clingo}{\texttt{clingo}~}

\newcommand{\titlep}[1]{ \smallskip \noindent \textbf{#1}}
\newcommand{\titleit}[1]{\smallskip \noindent \underline{\textit{#1.}}}

\newcommand{\maxes}{\mathsf{maxES}}
\newcommand{\maxec}{\mathsf{maxEC}}
\newcommand{\maxsc}{\mathsf{maxSC}}
\newcommand{\maxss}{\mathsf{maxSS}}
\newcommand{\minvc}{\mathsf{minAC}}
\newcommand{\minvpc}{\mathsf{minVC}}
\newcommand{\minvps}{\mathsf{minVS}}
\newcommand{\minvs}{\mathsf{minAS}}
\newcommand{\minv}{\mathsf{minA}}
\newcommand{\minvp}{\mathsf{minV}}
\newcommand{\maxs}{\mathsf{maxS}}
\newcommand{\maxe}{\mathsf{maxE}}

\newcommand{\SAT}{\textsc{3SAT}\xspace}
\newcommand{\fff}{\mathit{fff}}
\newcommand{\fft}{\mathit{fft}}
\newcommand{\ftf}{\mathit{ftf}}
\newcommand{\ftt}{\mathit{ftt}}
\newcommand{\tff}{\mathit{tff}}
\newcommand{\tft}{\mathit{tft}}
\newcommand{\ttf}{\mathit{ttf}}
\newcommand{\ttt}{\mathit{ttt}}

\newcommand{\HRO}{\Gamma_h^o}
\newcommand{\SRO}{\Gamma_s^o}
\def\hsrules{\Gamma}
\def\hsrulesO{\Gamma_{O}}
\def\hsrulesV{\Gamma_{V}}
\newcommand{\merO}{E}
\newcommand{\merV}{V}
\newcommand{\extdat}[3]{{#1}_{{#2},{#3}}}

\def\harrow{\Rightarrow}
\def\sarrow{\dashrightarrow}

\def\eqrel{\mathsf{EqRel}}

\newcommand{\np}{\textrm{NP}\xspace}
\newcommand{\conp}{\textrm{coNP}\xspace}
\newcommand{\PTIME}{\textrm{PTIME}\xspace}
\newcommand{\false}{\texttt{false}\xspace}
\newcommand{\true}{\texttt{true}\xspace}

\def\qvars{\mathsf{qvars}}
\def\qcons{\mathsf{cons}}
\def\dcons{\mathsf{Dom}}
\def\ratom{\pi}
\newcommand{\myi}{(\emph{i})\xspace}
\newcommand{\myii}{(\emph{ii})\xspace}
\newcommand{\myiii}{(\emph{iii})\xspace}

\newcommand{\spec}{\Sigma}
\newcommand{\merE}{E}
\newcommand{\dat}{\mathbb{D}}

\newcommand{\LOGSPACE}{\textsf{LOGSPACE}}

\newcommand{\rulesO}{\Gamma_O}

\newcommand{\denC}{\Delta}
\newcommand{\den}{\delta}
\newcommand{\schema}{\S}

\newtheorem{example}{Example}
\newtheorem{theorem}{Theorem}
\newtheorem{definition}{Definition}
\newtheorem{proposition}{Proposition}
\newtheorem{lemma}{Lemma}

\title{Advances in Logic-Based  Entity Resolution: \\Enhancing \textsc{ASPen} with Local Merges and Optimality Criteria}
\author{%
Zhiliang Xiang$^1$ \and  Meghyn Bienvenu$^{2}$ \and Gianluca Cima$^3$ \and\\[1mm] Víctor Gutiérrez-Basulto$^1$ \and Yazmín Ibáñez-García$^1$\\
\affiliations
$^1$Cardiff University, UK\\
$^2$Univ. Bordeaux, CNRS, Bordeaux INP, LaBRI, UMR 5800, Talence, France\\
$^3$Sapienza University of Rome, Italy\\
\emails
 \{xiangz6,gutierrezbasultov,ibanezgarciay\}@cardiff.ac.uk, 
meghyn.bienvenu@labri.fr,  cima@diag.uniroma1.it
}

\begin{document}

\maketitle

\begin{abstract}
 We present $\textsc{ASPen}^{\!+}$, which extends an existing ASP-based system, \textsc{ASPen}, for collective entity resolution with 
two important functionalities: support for local merges and new optimality criteria for preferred solutions. Indeed, 
\textsc{ASPen} only supports so-called global merges of entity-referring constants (e.g.\ author ids), in which all occurrences of matched constants are treated as equivalent and merged accordingly. 
However, it has been argued that when resolving data values, local merges are often more appropriate, as e.g.\ 
some instances of ‘J. Lee’ may refer to ‘Joy Lee’, while others should be matched with ‘Jake Lee’.
In addition to allowing such local merges, $\textsc{ASPen}^{\!+}$ offers new optimality criteria for selecting solutions,  
such as minimizing rule violations or maximizing the number of rules supporting a merge. 
Our main contributions are thus 
 (1) the formalization and computational analysis of various notions of optimal solution, and (2) 
an extensive experimental evaluation on real-world datasets, demonstrating the effect of local merges and the new optimality criteria on both accuracy and runtime.
\end{abstract}

\section{Introduction}

Entity Resolution (ER) is a foundational task in computer science, concerned with identifying and merging references (constants) that refer to the same real-world entity~\cite{DBLP:conf/kdd/GetoorM13}. A variety of ER approaches have been explored, differing in their assumptions, the characteristics of the data they handle, and the techniques they employ~\cite{ChristophidesACMSurvey21}. One general and expressive variant, known as \emph{collective ER}, involves the joint resolution of entity references across multiple interrelated tables~\cite{BhattacharyaTKDD07}. Declarative approaches have proven particularly well-suited for such complex, multi-relational settings, as they naturally leverage underlying relational dependencies~\cite{lace_2022,DengFLLZA22,FaginKL0S23,XiangBCGI24}. Nonetheless, and somewhat surprisingly, there are still relatively few logic-based systems that fully support collective ER~\cite{DengFLLZA22,XiangBCGI24}.  

One such system is \textsc{ASPen}~\cite{XiangBCGI24}, recently developed within the knowledge representation and reasoning community using answer set programming (ASP)~\cite{DBLP:journals/cacm/BrewkaET11,asp-in-prac-2012,DBLP:books/sp/Lifschitz19}. \textsc{ASPen} implements the \textsc{Lace} framework~\cite{lace_2022} for collective ER, a logical approach that supports recursive, global, explainable, and constraint-aware ER. \textsc{ASPen} builds on the foundational result that \LACE solutions can be faithfully represented using ASP stable models. It extends this theoretical insight by tackling key implementation challenges. Most notably, it addresses the efficient computation of similarity facts within rule bodies. As proposed in \textsc{Lace}, it computes not only individual ER solutions but also an entire space of set-maximal solutions, enabling the derivation of both possible and certain merges. To support scalable reasoning, \textsc{ASPen} approximates hard-to-compute merge sets (e.g., possible and certain merges) using lower- and upper-bound variants that retain formal guarantees. \textsc{ASPen} supports explanatory reasoning using xclingo~\cite{xclingo2}, offering justifications for each merge via proof trees, making it a justifiable ER system. Experimentally, \textsc{ASPen} demonstrates strong performance, particularly in multi-relational settings, outperforming existing rule-based systems like Magellan~\cite{magellan-2016} and JedAI~\cite{jedai-2020} in F1-score across several datasets. It effectively leverages recursion to uncover deeper merges, and its optimized similarity computation approach reduces memory usage by up to 99.6 \%, when compared to a naive method. Overall, despite higher computation times, \textsc{ASPen} proves competitive and scalable for complex, real-world ER settings.

The introduction of \textsc{ASPen}, along with the public release of its code and associated data, 
opens up new opportunities to further explore and evaluate ASP techniques in the context of ER. In this paper,  we introduce $\textbf{\textsc{ASPen}}^{\!+}$, which extends \textsc{ASPen}  with 
two important functionalities: support for local merges and new optimality criteria for preferred solutions. In the original \textsc{ASPen}, constant identification is global: all occurrences of matched constants are merged, not just those involved in the derivation of the match. This global semantics is particularly well suited for merging constants that serve as entity references, such as author names or paper IDs, and has been adopted in several logic-based frameworks~\cite{ArasuICDE09,BurdickTODS2016,DengFLLZA22,FaginKL0S23,lace_2022}. However, it has been argued that when resolving data values, local merges are often more appropriate~\cite{BertossiTCS13,FanPODS08,FanVLDB09,BienvenuCGI23,FaginKL0S23}. Indeed, a local semantics allows  
some instances of ‘J. Lee’ may refer to ‘Joy Lee’, while others should be matched with ‘Jake Lee’.   $\textsc{ASPen}^{\!+}$ also introduces additional optimality criteria for defining preferred solutions, extending beyond the approach used in \textsc{ASPen}, which prioritizes solutions with the greatest number of merges (\wrt set inclusion).  It supports seven optimality criteria, such as minimizing rule violations or maximizing the number of rules supporting a merge. Our main contributions are thus 
 (1) the formalization of various notions of optimal solution and computational analysis of the problem of recognizing optimal solutions \wrt the chosen optimality criteria, and (2) an extensive experimental evaluation on real-world datasets, demonstrating the effect of local merges and the new optimality criteria on both accuracy and runtime.

\smallskip 
\noindent
\textbf{Related Work} Several declarative systems for ER have been proposed, including Magellan~\cite{magellan-2016} and JedAI~\cite{jedai-2020}. However, most of these systems are designed for pairwise ER. 
Apart from \textsc{ASPen}, the most closely related logic-based ER approaches supporting multi-relational settings and recursion are Dedupalog~\cite{ArasuICDE09} and MRL~\cite{DengFLLZA22}. Nevertheless, these implementations are not publicly available and differ from $\textsc{ASPen}^{\!+}$ in key aspects.: i) $\textsc{ASPen}^{\!+}$ explores a space of preferred solutions under multiple optimality criteria, rather than producing a single solution; ii) it supports both local and global merges, unlike these systems which are limited to global merges only. Note that, like \textsc{ASPen}, $\textsc{ASPen}^{\!+}$ is an implementation of the \laceEXT framework~\cite{BienvenuCGI23}, which extends \textsc{Lace} to support local roles. In addition, $\textsc{ASPen}^{\!+}$ goes beyond \laceEXT by exploring a range of optimality criteria. We also note in passing that the framework proposed by~\citeauthor{FaginKL0S23}~\citeyear{FaginKL0S23}, like \laceEXT, supports both global and local merges, although no system currently implements it.

{Programs, code and data  of \ours are  available at~\cite{github}}

\newcommand{\const}{\mathbf{C}}
\newcommand{\objs}{\mathbf{O}}
\newcommand{\vals}{\mathbf{V}}
\newcommand{\tids}{\mathbf{TID}}
\newcommand{\vars}{\mathbf{X}}
\newcommand{\rels}{\mathbf{R}}
\newcommand{\typef}{\mathbf{type}}
\section{Preliminaries}
\subsubsection{Databases} 
We assume an infinite set of \emph{constants} $\const$ partitioned into three disjoint sets: a set of \emph{objects}, $\objs$, serving as references to real-world entities, a set $\vals$ of \emph{values}, and a set $\tids$ of tuple identifiers (tids), used to annotate database facts. 
A \emph{(database) schema $\mathcal{S}$} consists of a finite set of relation symbols $\rels$, each having an associated arity $k \in \mathbb{N}$ and type vector $\{\objs, \vals\}^k$.
We write $R/k \in \mathcal{S}$ to indicate that $R$ has arity $k$ and denote by $\typef(R,i)$ the $i$th element of $R$'s type vector. If $\typef(R,i) = \Obj$ (resp.\ $\Val$), we call $i$ an \emph{object (resp.\ value) position} of $R$.

 A \emph{$\TID$-annotated database  over a schema $\mathcal{S}$} is a finite set of facts $D$ of the form $R(c_0, c_1, \ldots, c_k)$, where $R/k \in \mathcal{S}$, $c_0 \in \tids$ and $c_i \in \objs \cup \vals$ for $1 \leq i \leq k$.   We require that each $t \in \tids$ occurs at most once in $D$. For simplicity, we refer to such a set of facts as a \emph{database}. We write $\ObjD(D)$ to denote the set of objects occurring in $D$ and  $\StrPos(D)$ to denote the set of values cells of $D$: $\{\tup{t,i} \mid R(t, c_1, \dots, c_k) \in D, \typef(R,i)= \Val\}$.
\subsubsection{Queries} A \emph{conjunctive query (CQ)} takes the form $q(\vec{x})= \exists \vec{y}.\varphi(\vec{x},\vec{y})$, 
 where $\vec{x}$ and $\vec{y}$  are disjoint tuples of variables, 
 and $\varphi(\vec{x}, \vec{y})$ is a conjunction of  
 relational atoms of the form $R(t_0,t_1, \ldots, t_k)$ where $R/k$ is a relation symbol and $t_i \in \const \cup \vec{x} \cup \vec{y}$ for $0 \leq i \leq k$, as well as similarity atoms.
\subsubsection*{Constraints}
 We will consider \emph{denial constraints} (DC) of the form $\forall \vec{x}. \neg \varphi(\vec{x})$, where $\varphi(\vec{x})$ is a conjunction of atoms which are either relational atoms or inequality atoms $t_1 \neq t_2$  with variables from $\vec{x}$. A database $D$ \emph{satisfies a DC} if and only if 
 $\exists \vec{x}.\varphi(\vec{x})$ is not satisfied in $D$. 
 Notably, DCs generalize the well-known class of functional dependencies (FDs).

\section{\laceEXT Entity Resolution Framework}\label{sec:framework}
In this section, we briefly recall \laceEXT 
and introduce new optimality criteria for selecting preferred ER solutions. Due to space constraints, the presentation 
focuses solely on the material relevant to \textsc{ASPen}$^{\!+}$. For a complete account, we refer the interested reader to~\cite{BienvenuCGI23}.


\subsection{ER Specifications}
Entity resolution is concerned with identifying pairs of syntactically distinct database constants that refer to the same thing (such pairs will be called \emph{merges}). 
In \laceEXT, 
a distinction is made between objects and values, and thus the attributes of database relations are typed as object or value attributes. 
Objects are references to real-world entities, e.g.\ paper and author identifiers, 
whereas values specify the properties of such entities, e.g.\ name and phone number of authors. 
Crucially, all occurrences of an object (possibly across different database tables) 
are assumed to refer to the same entity. This means that if objects $o$ and $o'$ are deemed to be the same, 
then we merge \emph{all} occurrences of $o$ and $o'$,  while the meaning of a value can depend on its context, 
so e.g.\ one occurrence of a value \emph{J. Lee} might be merged with \emph{Joy Lee}, while another could merge with \emph{Jack Lee} (or remain unmerged). 
For this reason, objects and values need to be handled differently during the ER process. 

\subsubsection*{Rules for Objects} We use \emph{hard} and \emph{soft} rules for objects to identify pairs of objects that must or might refer to the same real-world entity. 
Such rules have the following forms: 
$$
    q(x,y) \harrow \linkO(x,y) \qquad\qquad q(x,y) \sarrow \linkO(x,y)
$$ 
where $q(x,y)$ is a CQ 
whose free variables $x$ and $y$ occur only in object positions. 
Note that $\linkO$ is a special relation symbol 
used to store the merged pairs of objects. 
While hard and soft rules have essentially the same syntactic form (except for the type of arrow), they differ in meaning: hard rules indicate pairs of objects that \emph{must} be merged, whereas soft rules indicate likely merges. 


\subsubsection*{Rules for Values} 
We use \emph{hard and soft rules for values} 
to handle local identifications of values which are non-identical representations of the same information. Such rules 
take the following forms (using $\harrow$ for hard rules, $\sarrow$ for soft rules):
$$
    q(x_t,y_t) \rightarrow \linkV(\tup{x_t,i},\tup{y_t,j})  \quad \rightarrow \,\,\in \{\harrow, \sarrow\} 
$$
where $q(x_t,y_t)$ is a CQ 
whose variables $x_t$ and $y_t$ each occur once in $q$ in position $0$ of 
atoms with relations $R_x$ 
and $R_y$  
respectively, 
and $i$ and $j$ are value positions of $R_x$ and $R_y$. 
Here the special relation symbol $\linkV$ 
is used to store pairs of \emph{value cells} (i.e.\ tuple-position pairs) which have been merged. Intuitively, the above hard (resp.~soft) rule states that a pair of tids $(t_1,t_2)$ being an answer to $q$ provides sufficient (resp.~reasonable) evidence for concluding that the values in cells $\tup{x_t,i}$ and $\tup{y_t,i}$ are non-identical representations of the same information. 



\subsubsection*{Specifications} In \laceEXT, ER specifications consist of hard and soft rules for objects and values, together with a set of denial constraints that define what counts as a consistent database. 

\begin{definition}\label{def:ER}
    A \laceEXT \emph{entity resolution (ER) specification} $\E$ 
    takes the form $\E=\tup{\hsrulesO,\hsrulesV,\Delta}$, where $\hsrulesO = \HRO \cup \SRO$ is a finite set of hard and soft rules for objects, 
    $\hsrulesV=\HRV \cup \SRV$ is a finite set of hard and soft rules for values, 
    and $\Delta$ is a finite set of denial constraints. 
\end{definition}

\begin{example}
Figure~\ref{fig:EXSpec} shows a specification $\E$. The soft rule $\sigma_o^1$ suggests that author records with similar names and matching birth date and place (\dob, \pob) likely refer to the same author. The hard rule $\rho^1_v$ enforces merging similar names for identical author IDs, preserving the functional dependency 
from constraint $\delta_1$.
\end{example}
\begin{figure*}
\begin{subfigure}[t]{0.60\textwidth}
    \centering
    \begin{tabular}{|c|c|c|c|c|}
    \hline
    \multicolumn{5}{|c|}{\Author( $\tid$,  $\aid$, \name, \dob, \pob)} \\
    \hline
      \tid   & \aid & \name & \dob & \pob\\
      \hline
       $t_1$  & $a_1$ & A. Turing & 23/07/1912 & London\\
       $t_2$  & $a_2$ & Alan Turing & 23/07/1912 & London \\
       $t_3$  & $a_3$ & Clerk Maxwell & 13/06/1831 & Edinburgh \\
       \hline
    \end{tabular}
    \end{subfigure}
    \hfill
    \begin{subfigure}[t]{0.35 \textwidth}
        \begin{tabular}{|c|c|c|}
        \hline
        \multicolumn{3}{|c|}{\Awrd($\tid$, $\aid$, \awrd)}\\
        \hline
        \tid & \aid & \awrd \\ \hline
          $t_4$ & $a_1$ & Smith's Prize(1936)  \\
          $t_5$ & $a_2$ & Smith's Prize \\ 
          $t_6$ & $a_3$ & Smith's Prize \\ \hline
        \end{tabular}
        \vspace{4pt}
    \end{subfigure}
    \hfill
    \begin{subfigure}[t]{\textwidth}
      \begin{center}
      \begin{small}
       The similarity predicate $\approx$, is such that  $\text{A. Turing}   \approx \text{Alan Turing}$ and $ 
          \text{Smith's Prize}  \approx \text{Smith's Prize(1936)}$ 
        \end{small}
        \end{center}
    \caption{Database $D$ over the schema $\mathcal{S} = \{\Author/4 , \Awrd/2\}$}
    \end{subfigure}
    %
    \begin{subfigure}[t]{\textwidth}
  \begin{small}
   \begin{align*}
\delta^1 &= \forall x_t, y_t, y, n_1, n_2,  d_1, d_2, p_1,p_2. \neg(\Author(x_t, y, n_1, d_1, p_1) \land \Author(y_t, y, n_2, d_2, p_2) \land n_1 \neq n_2 ) \\
          \sigma^1_o &= \exists x_t, y_t, n_1, n_2, d, p. \Author(x_t, y_1, n_1, d,p ) \land \Author(y_t, y_2, n_2,d,p) \land n_1 \approx n_2 \sarrow \linkO(y_1, y_2)    \\
          \rho^1_v &=  \exists y, n_1, n_2, d_1,d_2, p_1, p_2. \Author(x_t, y, n_1, d_1, p_1) \land \Author(y_t, y, n_2, d_2, p_2) \land n_1 \approx n_2 \Rightarrow \linkV(\tup{x_t,2},\tup{y_t,2})  \\
           \sigma^2_v &= \exists y,z,w. \Awrd(x_t, y, z) \land \Awrd(y_t, y, w) \land z \approx w  \sarrow \linkV(\tup{x_t, 2}, \tup{y_t, 2})
    \end{align*}
    \end{small} 
 \caption{\laceEXT  ER specification $ \E = \tup{\hsrulesO, \hsrulesV,\Delta}$, with $\hsrulesO = \{\sigma^1_0 \}$ , $\hsrulesV = \{ \rho^1_v, \sigma^2_v\}$ and $ \Delta = \{\delta^1\} $}
\label{fig:EXSpec}
  \end{subfigure}
\hfill  
 \vspace{5pt}

\begin{subfigure}[t]{0.55\textwidth}
    \centering
        \begin{tabular}{|c|c|c|c|c|}
    \hline
    \multicolumn{5}{|c|}{\Author( $\tid$,  $\aid$, \name, \dob, \pob)} \\
    \hline
      \tid   & \aid & \name & \dob & \pob\\
      \hline
       $t_1$  & $\hat{o}_1$ & $\hat{v}_1$ & \{23/07/1912\} & \{London\}\\
       $t_2$  & $\hat{o}_1$ & $\hat{v}_1$ & \{23/07/1912\} & \{London\}\\
       $t_3$  & $\{a_3\}$ & \{Clerk Maxwell \}& \{13/06/1831\} & \{Edinburgh\}\\
       \hline
    \end{tabular}
    \end{subfigure}
    \hfill
    \begin{subfigure}{0.45\textwidth}
       \centering
        \begin{tabular}{|c|c|c|}
        \hline
        \multicolumn{3}{|c|}{\Awrd($\tid$, $\aid$, \awrd)}\\
        \hline
        \tid & \aid & \awrd \\ \hline
          $t_4$ & $\hat{o}_1$ & $\hat{v}_2$ \\
          $t_5$ & $\hat{o}_1$ & $\hat{v}_2$\\ 
          $t_6$ & $\{a_3\}$ & \{Smith's Prize\} \\ \hline
        \end{tabular}
    \end{subfigure}
    
      \vspace{4pt}
\begin{subfigure}[t]{\textwidth}  
 where $\hat{o}_1 = \{a_1, a_2\}$,  $\hat{v}_1 = \{\text{A.Turing}, \text{Alan Turing}\}$ and
   $\hat{v}_2 = \{\text{Smith's Prize}, \text{Smith's Prize(1936)}\}$
  \caption{Extended database $\extdat{D}{\merO}{\merV}$ induced by the merges $(a_1, a_2)\in E$, and $(\tup{t_1,2},\tup{t_2,2})$, $(\tup{t_3, 2},\tup{t_4,2}) \in V$
  }
\end{subfigure}
  \hfill  
  \caption{A solution to an ER specification in \laceEXT}
  \label{fig:R-example}
\end{figure*}

\subsection{ER Solutions}
In \laceEXT, a solution to an ER specification $\E$ and database $D$
takes the form of a pair $\tup{\merO,\merV}$, 
where $E$ is an equivalence relation over $\ObjD(D)$ 
and $V$ is an equivalence relation over $\StrPos(D)$. 
Solutions thus specify which pairs of objects and which pairs of value cells are deemed to be the same. 
Naturally, the rules and constraints of the specification impose requirements on $E$ and $V$. 
%
Every solution must be obtained by repeatedly applying the hard and soft rules for objects and values, choosing to stop when no further hard rule can be applied and the denial constraints are satisfied. 
Importantly, the database and equivalence relations are updated to reflect the already applied rules, 
which can lead to new rules being applicable and/or constraints becoming (un)satisfied. 

The formal definition of solution relies upon the notion of an \emph{extended database $\extdat{D}{\merO}{\merV}$
induced by} 
$\tup{\merO,\merV}$. Basically, every occurrence of an 
object $o$ in the original database $D$ is replaced by the set of objects $\{o' \mid (o,o') \in E\}$ 
and the value 
 in cell $\tup{t,i} \in \StrPos(D)$ is replaced 
by the set of values occurring in the merged cells $\{ \tup{t',i'} \mid (\tup{t,i}, \tup{t',i'}) \in \merV \}$
(the tuple identifiers in position $0$ are left untouched). The semantics of queries and constraints are then suitably adapted
to handle such extended databases, whose cells now contain sets of constants rather than single constants 
(we refer readers to \cite{BienvenuCGI23} for the details). We say that a pair of objects $(o,o')$ is \emph{active} in $\extdat{D}{\merO}{\merV}$
\wrt a (hard or soft) rule for objects $q(x,y) \rightarrow \linkO(x,y)$ 
just in the case that 
$(o,o')$ is an answer to $q(x,y)$ evaluated over $\extdat{D}{\merO}{\merV}$. We also say that $\extdat{D}{\merO}{\merV}$ \emph{satisfies} a (hard or soft) rule for objects $r=q(x,y) \rightarrow \linkO(x,y)$ when all the pairs \emph{active} in $\extdat{D}{\merO}{\merV}$
\wrt $r$ occur in $\merO$. We can 
define in a similar fashion what it means for a pair $(\tup{t,i},\tup{t',i'})$ to be active in $\extdat{D}{\merO}{\merV}$ \wrt a rule for values as well as when $\extdat{D}{\merO}{\merV}$ \emph{satisfies} a (hard or soft) rule for values. We write $\extdat{D}{\merO}{\merV} \models \Lambda$ when $\extdat{D}{\merO}{\merV}$ satisfies each $\lambda \in \Lambda$. 
Finally, we will  use the notation 
$\eqrel(P,S)$ for the smallest equivalence relation on $S$ that extends $P \subseteq S \times S$. 
We are now ready to recall the formal definition of solutions in \laceEXT.




    \begin{definition}\label{def:SOL}
        Given an ER specification $\E=\tup{\hsrulesO,\hsrulesV,\Delta}$ 
        and a 
        database $D$, we call 
        $\tup{\merO,\merV}$ a \emph{candidate solution for $(D,\E)$} if it satisfies one of the following three conditions: 
        \begin{itemize}
            \item $\merO=\eqrel(\emptyset,\ObjD(D))$ and $\merV=\eqrel(\emptyset,\StrPos(D))$; 
            \item $\merO=\eqrel(\merO' \cup \{(o,o')\},\ObjD(D))$,  where 
            $\tup{\merO',\merV}$ is a candidate solution
            for $(D,\E)$ and $(o,o')$ is active in $\extdat{D}{\merO}{\merV}$ \wrt some rule $r \in \hsrulesO$; 
            \item $V=\eqrel(\merV' \cup \{(\tup{t,i},\tup{t',i'})\},\StrPos(D))$, where
            $\tup{\merO,\merV'}$ is a candidate solution for $(D,\E)$ and $(\tup{t,i},\tup{t',i'})$ is active in $\extdat{D}{\merO}{\merV}$ \wrt some rule $r \in \hsrulesV$. 
        \end{itemize}
        A \emph{solution} for $(D,\E)$ is 
        a candidate solution $\tup{\merO,\merV}$ 
        such that (a) $\extdat{D}{\merO}{\merV} \models \HRO$, (b) $\extdat{D}{\merO}{\merV} \models \HRV$, and (c) $\extdat{D}{\merO}{\merV} \models \Delta$. We use $\SOL(D,\E)$ for the set of solutions for $(D,\E)$.
    \end{definition}
    \begin{example}
    Consider the database $D$ and specification $\E$ in Figure~\ref{fig:R-example}.  
    A trivial candidate solution is $\tup{\merO_0,\merV_0}$, with  $\merO_0=\eqrel(\emptyset,\ObjD(D))$ and $\merV_0=\eqrel(\emptyset,\StrPos(D))$. 
    Note that $(a_1, a_2)$ is active in $D_{\merO_0, \merV_0}$ \wrt $\sigma^1_o$. 
    Let, $\merO_1 = \eqrel(\merO_0 \cup \{(a_1, a_2)\},\ObjD(D))$.  However, this causes a violation of $\delta^1$ due to differing names for the same $\aid$ in $t_1$ and $t_2$. 
    Applying $\rho^1_v$  merges cells $\tup{t_1, 2}$ and  $\tup{t_2, 2}$, yielding 
    $\merV_1 = \eqrel(\merV_0 \cup \{(\tup{t_1, 2},  \tup{t_2, 2})\},\StrPos(D))$.  The resulting induced database satisfies $\delta^1$, but  $(\tup{t_4,2}, \tup{t_5, 2})$ is still active in $D_{E_1, V_1}$ \wrt $\sigma^2_v$,
    allowing the possibility to extend $\merV_1$ to  
     $\merV_2 = \eqrel(\merV_1 \cup {(\tup{t_4,2}, \tup{t_5,2})},\StrPos(D))$.
  {Note that, $\tup{\merO_0,\merV_0}$, $\tup{\merO_1,V_1}$ and $\tup{\merO_1,V_2}$ are all solutions for $(D,\E)$, as their corresponding induced databases satisfy all denial constraints and hard rules in the specification $\E$, whereas $\tup{\merO_1, \merV_0}$ is not, as it induced database violated $\delta^1$.}
    \end{example}

 \subsection{Optimality Criteria for ER Solutions}\label{sec:opt-crt}
In general,  a \laceEXT specification may give rise to zero, one, or many solutions. The absence of solutions results 
 from constraint violations that cannot be resolved through allowed merges, while multiple solutions arise from the choice 
 of which soft rule applications to perform (as the constraints may block some combinations of merges). 
 In the original work on \laceEXT, a notion of maximal solution was introduced to focus on 
solutions which contain the most merges (w.r.t.\ set inclusion). 
While 
this notion is quite reasonable, 
other natural optimality criteria can be used to select preferred solutions. Indeed, we may want to 
give more importance to a merge that is supported by multiple rules, 
or compare solutions based upon soft rule violations. 

To formalize these alternative criteria, we introduce the notation $\activepairs(D, E,V, \Gamma)$ for the set of all $(p,r)$ such that pair $p$ is 
active in $\extdat{D}{\merO}{\merV}$ w.r.t.\ rule $r \in \Gamma$. 
Our proposed optimality criteria are obtained by associating each 
solution $\tup{\merO,\merV}$ with one of the following sets
(using $\Gamma$ for $\hsrulesO \cup \hsrulesV$): 
\begin{align*}
\mergeset{(\merO,\merV)} &= \merO \cup \merV \\
\suppset{(\merO,\merV)} &= \{(p,r) \in \activepairs(D, E,V, \Gamma) \mid p \in \merO \cup \merV \}\\
\vioset{(\merO,\merV)} &= \{p \mid\! (p,r) \! \in \!\activepairs(D, E,V, \Gamma), p \not \in \merO \cup \merV\}\\ 
\viorset{(\merO,\merV)} &= \{(p,r) \in \activepairs(D, E,V, \Gamma) \mid p \not \in \merO \cup \merV \}
\end{align*}
then apply either set-inclusion or cardinality maximization or minimization. 
Observe that $\suppset{(\merO,\merV)}$ refines $\mergeset{(\merO,\merV)}$ by indicating the supporting rules for merges.
Likewise, $\vioset{(\merO,\merV)}$ gives only the active but absent pairs, while $\viorset{(\merO,\merV)}$ records which soft rules the absent pair violates. 
The resulting optimality criteria are as follows: 
\begin{itemize}
\item $\maxes$/$\maxec$: maximize $\mergeset{(\merO,\merV)}$
\item $\maxss$/$\maxsc$: maximize $\suppset{(\merO,\merV)}$
\item $\minvs$/$\minvc$: minimize $\vioset{(\merO,\merV)}$
\item $\minvps$/$\minvpc$: minimize $\viorset{(\merO,\merV)}$
\end{itemize}
where the final $\mathsf{S}$ (resp.\ $\mathsf{C}$) indicates comparison using set-inclusion (resp.\ set cardinality). 
For example, a solution $\tup{\merO,\merV}$ is $\minvpc$-optimal if there is no other solution $\tup{\merO',\merV'}$
such that $\viorset{(\merO',\merV')}<\viorset{(\merO,\merV)}$. Note that $\maxes$-optimal solutions correspond
to the maximal solutions of \cite{BienvenuCGI23}. {See Appendix~\ref{app:alt-opt} for further intuitions on the criteria.}
Interestingly, the optimality criteria give rise to different sets of optimal solutions, except for $\maxes$ and $\maxss$, which actually coincide. As a result, there are seven distinct optimality criteria overall.

\begin{proposition}\label{prop:rel}
    All pairs of defined criteria produce different sets of optimal solutions, except for $\maxes$ and $\maxss$.
\end{proposition}

%
\subsection{Recognizing Optimal Solutions}
A key challenge is determining if a solution is optimal under a given criterion. This problem is coNP-complete in data complexity for maxES (Bienvenu et al. 2023), and we show the same for the other criteria.

\begin{theorem}\label{conp-opt}
    For all seven optimality criteria, recognition of optimal solutions is $\conp$-complete in data complexity.
\end{theorem}
\begin{proof}[Proof sketch]
The upper bound is based on a guess-and-check procedure, analogous to the one in~\cite{BienvenuCGI23} for $\maxes$. We now give a reduction from the complement of $\SAT$. 
We use the following fixed schema:
\begin{align*}
\S=\{&R_{\fff}/3,~R_{\fft}/3,~R_{\ftf}/3,~R_{\ftt}/3,~R_{\tff}/3,~R_{\tft}/3,\\
&R_{\ttf}/3,~R_{\ttt}/3,~V/1,~F/1,~T/1,~B/1\}
\end{align*}
with only object attributes. 
Given an input $\varphi=c_1 \wedge \ldots \wedge c_m$ to $\SAT$ over 
variables $ x_1,\ldots,x_n$, where $c_i= \ell_{i,1} \vee \ell_{i,2} \vee \ell_{i,3}$, 
we construct an $\S$-database $D^{\varphi}$ that contains: 
    \begin{itemize}
        \item the fact $V(x_i)$ for each $i \in [1,n]$,
        \item the facts $T(1)$, $F(0)$, $B(0)$, and $B(1)$;
                \item for each clause $c_i=\ell_{i,1} \vee \ell_{i,2} \vee \ell_{i,3}$ in $\varphi$,
         the fact $R_{b_1 b_2 b_3}(x_{i,1}, x_{i,2}, x_{i,3})$ 
        with $\ell_{i,j} = (\neg) x_{i,j}$ and 
        $b_{i,j}=t$ (resp.\ $b_{i,j}=f$) if $\ell_{i,j}$ is a positive (resp.\ negative) literal
    \end{itemize}
We consider the fixed ER specification $\Sigma_{\SAT}=\tup{\hsrulesO,\emptyset,\Delta}$, 
where $\hsrulesO$ contains a single soft rule for objects: 
\begin{itemize}
\item $\sigma =V(x) \wedge B(y) \sarrow \linkO(x,y)$
\end{itemize}
and $\Delta$ contains the following denial constraints (all the variables are implicitly universally quantified):
 \begin{itemize}
            \item $\delta_0= \neg(F(y) \wedge T(y))$
            \item $\delta_1=\neg(R_{\fff}(y_1,y_2,y_3) \wedge T(y_1) \wedge T(y_2) \wedge T(y_3))$
            \item $\delta_2, \ldots, \delta_8$ defined analogously to $\delta_1$, but for relations $R_{\fft}$, $R_{\ftf}$, $R_{\ftt}$, $R_{\tff}$, $R_{\tft}$, $R_{\ttf}$, $R_{\ttt}$
            \item $\delta_9 =\neg(V(v) \wedge B(v) \wedge V(x) \wedge T(y) \wedge F(z) \wedge$ \\ $\phantom{d}\qquad\quad  x \neq y \wedge x \neq z)$
\end{itemize}          
The soft rule $\sigma$ allows a variable to merge with $0$ or $1$ (but not both due to $\delta_0$). Constraints $\delta_1$ to $\delta_8$ ensure that the chosen truth values 
do not violate any clause. 
Finally, constraint $\delta_9$ is used to ensure that if even one variable is merged with a truth value, using the sole soft rule $\sigma$, then \emph{every} variable must be merged with a truth constant. 
Together, the constraints ensure that the soft rule can only be applied if the formula is satisfiable. 
It can  thus 
be shown that 
$\varphi$ is unsatisfiable if and only if $\tup{E_\mathsf{triv}, \emptyset}$ is an $X$-optimal solution for $(D^{\varphi}, \Sigma_{\SAT})$, where $\merO_\mathsf{triv}=\eqrel(\emptyset,\ObjD(D^{\varphi}))$ and $X$ denotes any of the seven optimality criteria.
\end{proof}
Interestingly,~\cite{BienvenuCGI23} also studied the above problem in a \emph{restricted setting}, which forbids the presence of inequality atoms in denial constraints. It was shown that, under this restriction, the problem becomes tractable for $\maxes$, and in fact $\PTIME$-complete in data complexity. We now show that the same holds for all optimality criteria based on set inclusion, but not for those based on cardinality.

\begin{theorem}\label{restr}
    In the restricted setting, recognition of optimal solutions becomes $\PTIME$-complete in data complexity for the optimality criteria $\maxes$, $\minvs$, and $\minvps$, while it remains $\conp$-complete in data complexity for the optimality criteria $\maxec$, $\maxsc$, $\minvc$, and $\minvpc$.
\end{theorem}
\begin{proof}[Proof sketch] \textbf{$\PTIME$ cases.}
    We provide only a sketch of the upper bound for $\minvs$. Given $(D,\E)$ and $\tup{E,V}$, we check whether $\tup{E,V} \in \SOL(D,\E)$ (if not, then we return $\false$ and we are done). For each $p \in \vioset{(\merO,\merV)}$, we then check whether $E \cup V \cup \{p\}$ can lead to a solution for $(D,\E)$ having a strictly smaller set of active pairs than $\vioset{(\merO,\merV)}$. 

    To perform this latter check, we iteratively include all the new pairs that become active due to some hard rule (which may already occur in the first iteration due to the addition of $p$ to $E \cup V$) as well as all the new pairs that become active due to some soft rule and were not originally in $\vioset{(\merO,\merV)}$. Once a fixpoint is reached, we check whether the resulting $\tup{E',V'}$ is such that $\tup{E',V'} \in \SOL(D,\E)$, implying that $\tup{E',V'}$ is a solution such that $\vioset{(\merO',\merV')} \subsetneq \vioset{(\merO,\merV)}$. If for some $p \in \vioset{(\merO,\merV)}$ this is the case, then we return $\true$; otherwise, we return $\false$. The correctness of the above procedure is an immediate consequence of the following property which holds in the restricted setting: if $\extdat{D}{E'}{V'} \not\models \Delta$ and $E' \cup V' \subseteq E'' \cup V''$, then $\extdat{D}{E''}{V''} \not\models \Delta$ as well. In particular, this means that if a pair $\tup{E',V'}$ obtained as above is such that $\extdat{D}{E'}{V'} \not\models \Delta$, then it is not possible to obtain a solution for $(D,\E)$ by adding further merges.

    \textbf{$\conp$ cases.} We give a reduction from the complement of $\SAT$ for the optimality criteria based on cardinality. We use the fixed schema $\S'=\S \cup \{H/2\}$ with only object attributes, where $\S$ is as in the proof of Thm~\ref{conp-opt}. Given an input $\varphi$ to $\SAT$ over variables $x_1,\ldots,x_n$, we construct an $\S'$-database $\D^{\varphi}=D^{\varphi} \cup \{H(c_1,c_2)\}$, where $D^{\varphi}$ is as in the proof of Theorem~\ref{conp-opt} while $c_1$ and $c_2$ are two fresh constants. 
    
    We consider the fixed ER specification $\E'_{\SAT}=\tup{\hsrules,\Delta}$, where $\hsrules$ contains the soft rule $\sigma$ illustrated in the proof of Theorem~\ref{conp-opt} as well as the following additional soft rule: 
    \begin{itemize}
        \item $\sigma'=H(x,y) \sarrow \linkO(x,y)$
    \end{itemize}
    and $\Delta$ contains the following denial constraints (all the variables are implicitly universally quantified):
    \begin{itemize}
        \item $\delta_0= \neg(F(y) \wedge T(y))$
        \item $\delta_1=\neg(R_{\fff}(y_1,y_2,y_3) \wedge T(y_1) \wedge T(y_2) \wedge T(y_3) \wedge$ \\ $\phantom{d}\qquad\quad H(z,z))$
        \item $\delta_2, \ldots, \delta_8$ defined analogously to $\delta_1$, but for relations $R_{\fft}$, $R_{\ftf}$, $R_{\ftt}$, $R_{\tff}$, $R_{\tft}$, $R_{\ttf}$, $R_{\ttt}$
    \end{itemize}
    Essentially, the denial constraints are similar to those in the proof of Theorem~\ref{conp-opt}, except that $\delta_1,\ldots,\delta_8$ also contain the conjunct $H(z,z)$. This latter conjunct holds if and only if $c_1$ merges with $c_2$, which is possible due to $\sigma'$. 
    
    Now, let $E^0=\{(x_i,0) \mid 1 \leq i \leq n\}$ and $E=\eqrel(E^0,\ObjD(\D^{\varphi}))$. Note that $\tup{E,\emptyset} \in \SOL(\D^{\varphi},\E'_{\SAT})$ (because $(c_1,c_2) \not \in E$) and $\vioset(E,\emptyset)=\{(x_i,1) \mid 1 \leq i \leq n\} \cup \{(c_1,c_2)\}$ (so, $|\vioset(E,\emptyset)|=n+1$). The only way to get a solution $\tup{E',\emptyset}$ such that $|\vioset(E',\emptyset)|<n+1$ is to merge $c_1$ and $c_2$ as well as all the variables with $0$ or $1$, which is possible if and only if $\varphi$ is satisfiable. It can thus be shown that $\varphi$ is unsatisfiable if and only if $\tup{E,\emptyset}$ is a $\minvc$-optimal solution for $(\D^{\varphi},\E'_{\SAT})$. Similar considerations apply for $\minvpc$, while for $\maxsc$ and $\maxec$ we need to introduce a copy of each variable into a new unary predicate, an additional soft rule allowing to merge such copies with $0$ or $1$, and an additional denial constraints specifying that no variable and its copy are assigned to the same truth value. This guarantees that each $\maxec$-optimal (resp.~$\maxsc$-optimal) solution has the same cardinality.
\end{proof}

%

\section{ASP Encoding and Implementation}
We assume basic familiarity with ASP, see~\cite{asp-in-prac-2012,DBLP:books/sp/Lifschitz19} for a detailed introduction. For our purposes, it suffices to focus on two constructs: \emph{normal rules}, which have a single atom in the head, and \emph{constraints}, which are rules with an empty head. 
We  denote an ASP program (a set of such rules) by $\Pi$. Given a program $\Pi$, 
its  \emph{grounding} is the set of all ground instances of rules in $\Pi$, using the constants that appear in $\Pi$, and $\ASet{\Pi}$ denotes the set of all its stable models.
In addition to solving the  decision problem of checking whether $\ASet{\Pi}$  is non-empty, ASP-solvers also support advanced reasoning features such as enumerating the first $n$ stable models of $\Pi$, projecting models onto a specified set of atoms, and enumerating the first $n$ projected models~\cite{DBLP:conf/ijcai/GebserLMPRS18}.
ASP solvers also support optimization, allowing the computation (or enumeration) of $n$ elements of $\ASet{\Pi}$ that optimize a specified objective. For example, the directive $\mathtt{\#minimize}\{w,\mathbf{t}:\mathbf{L}\}$ instructs the solver to minimize the weighted occurrence of tuples $\mathbf{t}$ with weight $w$, subject to a list of literals $\mathbf{L}$. 
\subsection{ASP Encoding of Solutions}~\label{sec:asp-enc}
Given an ER specification $\Sigma$ and a database $D$, we define an ASP program  \aspenc 
containing all the facts in $D$, and an ASP rule for each (hard or soft) rule in $\Sigma$.  
Consider, for example,  the specification $\Sigma^\prime$ in Figure~\ref{fig:R-example}. Rules are translated as follows: 
\begin{small}
\begin{equation*}
\begin{aligned}
     &\begin{aligned}
 \sigma^1_o :\{&\mathtt{eqo}(X,Y);\aspneqo(X,Y)\}\mathtt{=1} \leftarrow  \aspauthor(T, X, N, D, P),
      \\[-4pt]
      & \aspauthor(T^\prime, Y, N^\prime, D, P),  \mathtt{val}(T,2,N_1), \mathtt{val}(T^\prime,2,N_2),
      \\[-4pt]
      &\simpred(N_1,N_2,S), S\geq 95.
       \end{aligned}
       \\
   & \begin{aligned}
\rho^1_v :\aspeqv& (T,2,T^\prime,2)\leftarrow  
    \aspeqo(X,Y), \simpred(N_1,N_2,S), S\geq 95, 
    \\[-4pt]
      & \aspauthor(T, X, N, D, P),   \aspauthor(T^\prime, Y, N^\prime, D^\prime, P^\prime),
      \\[-4pt]
       & \mathtt{val}(T,2,N_1),  \mathtt{val}(T^\prime,2,N_2) .
       \end{aligned}
\\
   & \begin{aligned}
      \sigma^2_v:\{&\aspeqv(T,2,T^\prime,2);\aspneqv(T,2,T^\prime,2)\}\mathtt{=1}\leftarrow  
      \aspeqo(X,Y),
      \\[-4pt]
      &  \aspawarded(T, X, A),  \aspawarded(T^\prime, Y, A^\prime),  \mathtt{val}(T,2,A_1),
      \\[-4pt]
    & \mathtt{val}(T^\prime,2,A_2), \simpred(A_1,A_2,S), S\geq 95.
       \end{aligned}
       \\
    &\begin{aligned}
    \delta^1 :   \bot& \leftarrow \aspauthor(T, X, N, D, P),   \aspauthor(T^\prime, Y, N^\prime, D^\prime, P^\prime), 
    \\[-4pt]
    & \aspeqo(X,Y), \mathtt{not} \text{ } \mathtt{i}(T,2,T^\prime,2).
\end{aligned}
\end{aligned}
\end{equation*}
\end{small}
In brief, each relational atom in the body of a rule is translated into an atom in the body of an ASP rule. The relations $\aspeqo$ and $\aspeqv$ represent mandatory object and value merges, respectively, thereby encoding solutions to $(D, \Sigma)$. 
Atoms of the form $\aspeqo(X,Y)$ in ASP rule bodies are used to encode that instantiations of $X$ and $Y$ are determined to denote the same object.  To capture local semantics,  the program includes facts of the form $\mathtt{proj}(T,i,V)$, indicating that the $i$-th position of tuple $T$ has value $V$. Each $\mathtt{proj}$ fact is then assigned to a value set, identified by a cell, through local merges. This is formalized by the rule:
$\mathtt{val}(T,I,V) \leftarrow \mathtt{proj}(T,I,V), \aspeqv(T^\prime,J,T,I).$ Importantly, similarity measures over value positions are evaluated not on the original variables in the relational atoms, but on pairs of fresh variables associated with $\mathtt{val}$-atoms.
The program also includes rules to ensure that both $\aspeqo$ and $\aspeqv$ form equivalence relations. Soft constraints are implemented using choice rules, allowing the inclusion of a pair $(X,Y)$ (or $(T,2,T',2)$, respectively) in either $\aspeqo$ (or $\aspeqv$) or their respective negations $\aspneqo$ and $\aspneqv$.
Note that, in addition to the encoding format shown above, database atoms (facts) can also be represented in a reified format (see Appendix~\ref{app:reified}).

\titlep{Practical changes}
Following \textsc{ASPen}, we represent similarity between constants using a ternary relation $\mathtt{sim}_\mathtt{i}(X, Y, S)$, where 
$X$ and $Y$ are the constants being compared, and $S$ is the similarity score.
Null values are represented using the atom $\emptypred(nan)$, in which, also as in~\textsc{ASPen}, the special constant $nan$ is prevented from participating in any $\aspeqo$ (resp. $\aspneqo$) and $\simpred$ atoms. Cells with null values are allowed to participate in $\aspeqv$. However, when equality between value positions is checked in a rule body, e.g., using $\mathtt{val}(T,i,V), \mathtt{val}(T,j,V)$, we include the condition $\defneg~\emptypred(V)$ to exclude nulls from the comparison. \textcolor{black}{This encoding blocks merges that would arise from treating two nulls as equivalent in rule bodies.}  \textcolor{black}{Moreover, to capture the local semantics of inequalities in DCs, the encoding must ensure that the value sets of the two cells are disjoint.}
This requirement was overlooked  in~\cite{BienvenuCGI23}.
\textcolor{black}{To this end,} for each pair of variables $v$ and $v^\prime$ associated with cells $\tup{t,i}$ and $\tup{t^\prime,j}$, we add the following rule to capture value overlap: 
\begin{equation*}
\footnotesize
\mathtt{i}(T,i,T^\prime,j) \leftarrow \mathtt{val}(T,i,V), \mathtt{val}(T^\prime,j,V),  \defneg~\emptypred(V).
\end{equation*}
This rule asserts that the predicate $\mathtt{i}(T,i,T',j)$ holds if the two cells share a non-null value—that is, their value sets intersect. Consequently, inequality between the two positions can be expressed by the absence of such an intersection: $\defneg~\mathtt{i}(T,i,T',j)$ like in $\delta^1$.



\subsection{Computing Optimal Solutions}
%

To compute solutions under the $\maxsc$ criterion, we extend the predicates $\aspeqo$ and $\aspeqv$ (or $\aspneqo$ and $\aspneqv$) by adding an extra argument to represent the rule label. Each hard or soft rule in \aspenc is assigned a unique label constant, which is included in the head atom to distinguish merges derived from different rules. In rule bodies, any occurrence of $\aspeqo$ or $\aspeqv$ uses an anonymous variable in the label position to avoid hard-coding the label.
Finally, for each rule labeled $r$ that applies to objects (respectively, to values), we add a corresponding rule to \aspenc: $\mathtt{w}(X,Y,l,w) \leftarrow~\aspeqo(X,Y,r).$  (resp. $\mathtt{w}(T,i,T^\prime,j,r,w) \leftarrow~\aspeqv(T,i,T^\prime,j,r).$). This assigns a weight $w$ to merges derived from rule $r$.



The transformation for encoding rule violations under the $\minvp$ criterion follows a similar approach. Specifically, the head atoms of hard rules are annotated with a default label constant $r^\prime$, while those of soft rules are assigned distinct labels. \textcolor{black}{ For example, in~\sect{sec:asp-enc}, the head of the soft rule $\sigma^1_o$  is replaced by $\{\aspeqo(X,Y,\sigma^1_o);\aspneqo(X,Y,\sigma^1_o)\}\mathtt{=1}$, similarly for $\sigma^2_v$. } In contrast,  the head of the hard rule $\rho^1_v$ is expanded with the default label $r^\prime$. Moreover, rules that aggregate weights (e.g.\  those required for computing $\maxsc$-optimal solutions) will not be included.

\titlep{Set-based optimisation}
Set-based optimal solutions correspond to the stable models of $\Pi(D,\Sigma)$ that contain an optimal set of target facts, depending on the chosen criterion: (1) a maximal set of $\aspeqo$/$\aspeqv$ facts for $\maxe$; (2) a minimal set of (labeled) $\aspneqo$/$\aspneqv$ facts for $\minvp$ and $\minv$; (3) a maximal set of weighted facts for $\maxs$.
Following~\cite{XiangBCGI24}, we utilise the ASP-based preference framework asprin~\cite{DBLP:journals/ai/BrewkaDRS23} to encode preferences over the stable models of a program. For our uses, we prefer a model $M^\prime$ over $M$ if its projection on the target facts is a strict superset or subset of that of $M$, under the selected criterion.

Alternatively, domain-specific heuristics can be used to simulate set preferences by prioritizing 
the truth assignments of target facts 
\cite{RosaGM10,GebserKROSW13}. We encode such heuristics using the statement
$\mathtt{\#heuristic\ t.\ [1,\ p]}$,
where $\mathtt{t}$ is a non-ground atom with the same predicate as the target facts for the chosen criterion, and $\mathtt{p} \in \{\mathtt{True}, \mathtt{False}\}$ enables maximization or minimization, respectively.

\titlep{Cardinality-based optimisation}
To compute cardinality-optimal solutions, we use the standard ASP optimization statement~\cite{asp-in-prac-2012}, which minimizes (or maximizes) over the full tuple of arguments in the target atoms, using a default weight and priority of 1. 
For instance, to compute $\minvpc$, we include the following statements:
$\mathtt{\#minimize\{1@1,X,Y,R : neqo(X,Y,R)\}}$ and $\mathtt{\#minimize\{1@1,T,I,T',J,R : neqv(T,I,T',J,R)\}}.$
Alternatively, preferences over the cardinality of target facts can be expressed using asprin, allowing the selection of models with either maximal or minimal cardinality depending on the chosen criterion.
It is worth noting that while heuristic-driven solving and multi-threaded solving~\cite{GebserKS12} can both improve efficiency for optimization tasks, they are not compatible and cannot be used simultaneously.

\subsection{Implementation}

%
The workflow of \ours 
begins with an ASP encoding of an ER specification, a \texttt{CSV}/\texttt{TSV} dataset containing duplicate and potentially corrupted values (e.g., nulls), and a set of command-line options. A preprocessing component computes similarity scores between relevant constant pairs based on attributes referenced in similarity atoms. 
These scores, along with the dataset, are passed to a program transformer, which rewrites the encoding into the \ours format and generates ASP facts for both database tuples and similarities. Finally, the ER controller, built on \texttt{clingo}, computes optimal solutions according to the specified input options. See Appendix C for a workflow illustration.

We implemented \ours by extending the open-source system \prev. Specifically, we developed a new program transformer, based on the translation procedure from~\cite{BienvenuCGI23}, which enables users to write ASP rules in the style of ER specifications without dealing with the underlying complexities of ASP encoding. Additionally, we enhanced the ER controller to support computing solutions under various optimality criteria, leveraging ASP features such as heuristic-driven solving.



\section{Experiments}
\begin{table*}[t]
\renewcommand\arraystretch{0.15}
\setlength{\tabcolsep}{0.28em}
\centering
\begin{tabular*}{\linewidth}{@{}ccccc|ccc|c|cccc|ccc|c@{}}
\toprule
 \textbf{Method}  & \textbf{Data} &  \multicolumn{3}{c|}{$\fone$~~~~~(\textbf{P}~~  
 /~~~\textbf{R})} &  \textbf{$\ter$} & $\tground$ & $\tsolve$ & \textbf{\#DC}&\textbf{Data} &  \multicolumn{3}{c|}{$\fone$~~~~~(\textbf{P}~~  
 /~~~\textbf{R})} & \textbf{$\ter$} & $\tground$ & $\tsolve$  & \textbf{\#DC}\\

\midrule
Magellan & \multirow{4}{*}[-1.5ex]{\rotatebox{90}{\imdb}}  &
 88.09 & \textbf{99.80} & 78.83 & {\textbf{3.89}} & N/A & N/A  & 0/5 & \multirow{4}{*}[-1.5ex]{\rotatebox{90}{\corimdb}} 
 & 83.05& \textbf{99.77} & 71.12 & \textbf{3.11} & N/A & N/A  & 0/5 \\
 \cmidrule{1-1} \cmidrule{3-9} \cmidrule{11-17}
  JedAI & & 97.49& \underline{99.40} & 95.67 & \underline{18.78}  & N/A & N/A  & 0/5 &  & \underline{97.49}&  \underline{99.40} & \underline{95.67} & \underline{18.67}  & N/A & N/A   & 0/5 \\
 \cmidrule{1-1} \cmidrule{3-9} \cmidrule{11-17}
\prev & &\textbf{99.27} &99.39 & \textbf{99.14} & 610.49 &  \textbf{11.51} & \textbf{0.082}  & 1/5 & & 96.99 & 99.36 & 94.73 & 757.17 & \textbf{11.75} & \textbf{0.088}  & 1/5 \\
 \cmidrule{1-1} \cmidrule{3-9} \cmidrule{11-17}
  \ours & &\textbf{99.27} &99.39 & \textbf{99.14}  & 86.24 & 13.02 & \underline{0.096}  & 3/5 & & \textbf{99.13} & 99.39 & \textbf{98.87} & 94.85 & 18.71 & 0.69  & 3/5 \\
\midrule
\midrule
Magellan  & \multirow{4}{*}[-2.5ex]{\rotatebox{90}{\music}} & \underline{89.78} & \underline{98.63} & 82.38 & \textbf{64.83} & N/A & N/A  & 0/23 &\multirow{4}{*}[-2.5ex]{\rotatebox{90}{\cormusic}} & 55.54 & \textbf{97.51} & \underline{38.83} & \textbf{66.87} & N/A & N/A  &0/23\\
 \cmidrule{1-1} \cmidrule{3-9} \cmidrule{11-17}
JedAI & & 70.67 &87.46 & \underline{59.30} & \underline{105.06} &N/A & N/A  & 0/23 &  & 32.75 &73.95& 21.02 & \underline{101.02} & N/A & N/A  & 0/23
\\
 \cmidrule{1-1} \cmidrule{3-9} \cmidrule{11-17}
 \prev & & \textbf{97.64} & \textbf{99.45} & \textbf{95.89} & 666.01 &  \textbf{1.78} &  \textbf{0.20}  & 21/23 & & \underline{90.31} & 91.86  & \textbf{88.81} & 695.76 & \textbf{20.15} & \textbf{3.57}   & 17/23\\
 \cmidrule{1-1} \cmidrule{3-9} \cmidrule{11-17}
 \ours & &\textbf{97.64}& \textbf{99.45} & \textbf{95.89} & 665.65 & 1.82 & 0.21   & 23/23 &  & \textbf{90.46} & \underline{92.17} & \textbf{88.81} & 770.44 & 72.81 & 12.16   & 23/23\\
\midrule
\midrule
 Magellan & \multirow{4}{*}[-1.5ex]{\rotatebox{90}{\cellmusic}} &   55.48& \textbf{97.62} & 38.75 & \textbf{64.81} & N/A & N/A  & 0/23 &\multirow{4}{*}[-2.5ex]{\rotatebox{90}{\poke}} & 7.01& 3.97 & 29.74 & {\underline{260.96}} & N/A & N/A  & 0/10\\
 \cmidrule{1-1} \cmidrule{3-9} \cmidrule{11-17}
  JedAI & & 31.04& 72.47 & 19.75 & \underline{101.47} &N/A & N/A  & 0/23 & &  2.1& 1.08 & 46.56 & \textbf{23.46} & N/A & N/A  & 0/10\\
\cmidrule{1-1} \cmidrule{3-9} \cmidrule{11-17}
\prev & & \underline{71.18} & 77.83 & \underline{65.58} & 718.38 & \textbf{21.01} & \textbf{10.47}  & 16/23 & & \underline{81.78} & \underline{99.71}  & \underline{69.31} & 4,454 & \textbf{311.37} & \textbf{0.91}   & 10/10\\
 \cmidrule{1-1} \cmidrule{3-9} \cmidrule{11-17}
 \ours & & \textbf{88.85} & \underline{89.5}   & \textbf{88.21} & 993.31 & 97.51 & 23.02  & 23/23 &  & \textbf{84.98} & \textbf{99.73} & \textbf{74.03} & 11,880 & 496.14 & 48.52  & 10/10\\
\bottomrule
\end{tabular*}
\caption{Results on Complex Multi-relational Datasets}
\label{tbl:mrer}
\end{table*}
Our experiments focus on the following  questions:
(1) How does \ours perform compared to \textsc{SoTA} systems that rely exclusively on global semantics?
(2) What are the most efficient configurations for each optimization criterion? How do the resulting solutions vary in terms of performance? Can these solutions be enumerated efficiently?  We also present qualitative studies that highlight the value of integrating both global and local semantics to  address complex ER scenarios.

\smallskip
\titlep{Datasets and Metrics} 
%
We evaluate \ours on six multi-relational datasets: (i)~\imdb, based on the IMDB database, which includes entities such as artists and movies, and contains real-world duplicates; (ii)~\music, which features synthetic duplicates derived from the MusicBrainz database; (iii)~\cormusic, a noisier version of \music that retains the same duplicates but includes more missing values and syntactic variations; (iv)~\poke,  a larger-scale dataset from the Pokémon database, characterized by a greater number of tuples and more complex inter-table relationships. We also consider two synthetic datasets, \corimdb and \cellmusic. 
These are derived from \imdb and \cormusic, respectively, and contain a higher proportion of syntactic variants. Both datasets follow the variant generation protocol described in~\cite{XiangBCGI24}. We 
use the standard accuracy  metrics~\cite{DBLP-10} of \emph{Precision} (\textbf{P}), \emph{Recall} (\textbf{R}), and \emph{F1-Score} ($\fone$). Sources and statistics for all datasets, as well as details of the experimental environment and similarity measures, are provided in Appendix~\ref{app:setup}.




\subsection{Combining Global and Local Semantics}

\titlep{Setup} 
We evaluate the performance of \ours against three baselines: \prev and two pairwise rule-based ER systems, Magellan~\cite{magellan-2016} and JedAI~\cite{jedai-2020}. Both \ours and \prev use the same global rules from~\prev, with \ours additionally incorporating local rules. Functional dependencies (FD) from the dataset schema are included in both ASP-based systems, provided they yield satisfiable ER programs. For Magellan and JedAI, we apply the same preconditions and similarity measures as in the ASP-based systems to ensure comparability. Like in \textsc{ASPen}, we consider a single $\maxes$-solution as default output of~\ours. We report accuracy metrics  and total running time $\ter$ for all systems. For the ASP-based systems, we also report grounding time ($\tground$), solving time ($\tsolve$), and the number of DCs (\textbf{\#DC}) added without violating constraints. Results are shown in \tbl{tbl:mrer}.


\titlep{Accuracy}
%
%
\ours consistently achieved the highest F1 scores across all datasets. Compared to the pairwise baselines, it outperformed \magellan and \jedai by wide margins, with F1 gains ranging from 7.86\% to 82\% over \magellan, and 1.64\% to 77\% over \jedai. Similar trends were observed for \prev, highlighting the advantage of logic-based approaches that support collective ER and enforce constraints.

When compared to \prev, \ours matched its performance on \imdb and \music, but achieved F1 improvements of 2.14\%, 0.15\%, 17.6\%, and 3.2\% on \corimdb, \cormusic, \cellmusic, and \poke, respectively. Except for \imdb and \corimdb (where inherent constraint violations exist) \ours successfully incorporated all FDs from the schema. In contrast, \prev’s reliance on global semantics alone often led to unsatisfiable programs when all FDs were included. Thus, \ours produced solutions with better precision and recall, particularly on noisy datasets like \cellmusic. Interestingly, even in \poke, where both systems could include all FDs, \ours achieved 4.72\% higher recall than \prev, demonstrating that without handling alternative value representations, valid merges can still be blocked by DCs.

\titlep{Running time}
The overall time $\ter$ consists of preprocessing (similarity computation) and ER time.  
For ASP-based systems, the ER time is further composed of grounding and solving time.
\prev uses an optimized similarity algorithm for preprocessing, making similarity computation faster on more complex datasets. In contrast, \ours uses a naive cross-product similarity computation, as \textsc{ASPen}'s optimized algorithm is incompatible with local semantics.
\\
Overall, both \magellan~and \jedai significantly outperformed the ASP-based systems on $\ter$. Compared to \textsc{ASPen}$^{\!+}$, \magellan was faster by factors of 22, 30, 10, 11, 15, and 45 on the \imdb, \corimdb, \music, \corimdb, \cellmusic, and \poke, respectively. Similarly, \jedai achieved speed advantages of 4.5 to 
506 times over the same datasets. Due to differences in the similarity computation, \ours was 7 and 8 times faster than \prev on simpler datasets like \imdb and \corimdb. However, on more complex datasets, the optimised similarity algorithm used by \prev gave speed advantages of 1.1, 1.4, and 2.6 times compared to \ours on \cormusic, \cellmusic, and \poke.
\\
In terms of $\tground$ and $\tsolve$, \prev consistently outperformed \ours. On \cormusic, \cellmusic, and \poke, \prev’s grounding was 3.6, 4.6, and 1.5 times faster, respectively. The differences in solving time were even more pronounced: \ours was 2.1, 3.4, 7.8, and 53 times slower than \prev on \cellmusic, \cormusic, \corimdb, and \poke. These results show that while \ours delivers higher ER quality, it often does so at the expense of computational efficiency.

\subsection{Comparison on Optimization Criteria}
\textbf{Setup} For each dataset, we fixed the corresponding ER program and used \ours to compute one and up to 50 optimal solutions based on the set of optimization criteria defined in Section~\ref{sec:opt-crt}. For set-based criteria, heuristic-driven solving was enabled. For cardinality-based criteria, we used weighted constraints and enabled parallel solving with 36 threads.
We recorded the average accuracy metrics over the number of enumerated solutions $\mathbf{\#e}$ as $\overline{F}_1$/$\overline{P}$/$\overline{R}$. For efficiency, we measured the solving time to compute the first optimal model ($t^1_s$) and the average solving time per model $t^n_s$. For simplicity, we use the prefixes \emph{S-} and \emph{C-} to denote, in general, set-based and cardinality-based optimality criteria, respectively. Results are summarized in~\tbl{tbl:opt-s} and~\tbl{tbl:opt-c}.  Appendix~\ref{app:bench} presents results for various configurations under each optimality criterion (e.g., using aspirin vs. using heuristic). The results reported in this section correspond to the most efficient configuration.

\begin{table}[htbp]
\renewcommand\arraystretch{0.15}
\setlength{\tabcolsep}{0.28em}
\centering
\begin{tabular}{cc|ccc|ccc}
\toprule
\textbf{Data} & \textbf{Method} & \multicolumn{3}{c|}{$\avgfone$~~~~~($\avgp$~~/~~~$\avgr$)} & $\tenumone$ & $\enumc$ & $\tenumn$ \\
\midrule
\multirow{4}{*}[-0.5ex]{\rotatebox{90}{\cormusic}} 
& $\maxes/\mathsf{SS}$ & 90.50 & 92.20 & \textbf{88.86} & \textbf{12.16} & 50 & 1.86 \\
\cmidrule{2-8}
& $\minvs$ & \textbf{91.88} & \textbf{95.11} & \textbf{88.86} & 12.8 & 50 & \textbf{1.7} \\
\cmidrule{2-8}
& $\minvps$ & \underline{91.7} & \underline{94.73} & \underline{88.85} & \underline{12.44} & 50 & \underline{1.78} \\
\midrule
\multirow{4}{*}[0.5ex]{\rotatebox{90}{\cellmusic}} 
& $\maxes/\mathsf{SS}$ & 88.85 & 89.5 & \textbf{88.21} & 23.02 & 50 & \underline{1.67} \\
\cmidrule{2-8}
& $\minvs$ & \underline{89.62} & \underline{91.11} & 88.18 & 23.01 & 50 & \textbf{1.54} \\
\cmidrule{2-8}
& $\minvps$ & \textbf{90.13} & \textbf{92.2} & 88.15 & \textbf{21.61} & 50 & 2.48 \\
\midrule
\multirow{4}{*}[-0.5ex]{\rotatebox{90}{\poke}} 
& $\maxes/\mathsf{SS}$ & \textbf{84.98} & \textbf{99.73} & \textbf{74.03} & \textbf{48.52} & 1 & N/A \\
\cmidrule{2-8}
& $\minvs$ & 83.83 & \textbf{99.73} & 72.3 & 51.58 & 50 & \underline{0.06} \\
\cmidrule{2-8}
& $\minvps$ & \underline{84.62} & \textbf{99.73} & \underline{73.48} & 56.87 & 50 & \textbf{0.01} \\
\bottomrule
\end{tabular}
\caption{Result on optimal solutions under different set-optimisation criteria over the datasets}
\label{tbl:opt-s}
\end{table}

\begin{table}[htbp]
\renewcommand\arraystretch{0.15}
\setlength{\tabcolsep}{0.32em}
\centering
\begin{tabular}{cc|ccc|ccc}
\toprule
\textbf{Data} & \textbf{Method} & \multicolumn{3}{c|}{$\avgfone$~~~~~($\avgp$~~/~~~$\avgr$)} & $\tenumone$ & $\enumc$ & $\tenumn$ \\
\midrule
\multirow{4}{*}[-1.5ex]{\rotatebox{90}{\cormusic}} &
$\maxec$ & 90.51 & 92.20 & \textbf{88.9} & \underline{35.09} & 50 & \textbf{2.53} \\
\cmidrule{2-8}
& $\maxsc$ & 90.52 & 92.21 & \textbf{88.9} & \textbf{30.1} & 16 & 12.09 \\
\cmidrule{2-8}
& $\minvc$ & \underline{91.4} & \underline{94.05} & \textbf{88.9} & 84.69 & 50 & \underline{2.61} \\
\cmidrule{2-8}
& $\minvpc$ & \textbf{92.01} & \textbf{95.35} & \underline{88.89} & 48.97 & 50 & 5.66 \\
\midrule
\multirow{4}{*}[-0.5ex]{\rotatebox{90}{\cellmusic}} 
& $\maxec$ & 88.83 & 89.45 & \textbf{88.21} & 66.71 & 1 & N/A \\
\cmidrule{2-8}
&  $\maxsc$ & 88.85 & 89.50 & \textbf{88.21} & \textbf{52.8} & 2 & 629.56 \\
\cmidrule{2-8}
& $\minvc$ & \underline{89.51} & \underline{90.93} & 88.14 & 92.97 & 50 & \textbf{2.12} \\
\cmidrule{2-8}
& $\minvpc$ & \textbf{89.74} & \textbf{91.36} & \underline{88.18} & \underline{67.82} & 50 & \underline{8.42} \\
\midrule
\multirow{4}{*}[-1.5ex]{\rotatebox{90}{\poke}} 
& $\maxec$ & \textbf{84.98} & \textbf{99.73} & \textbf{74.03} & \textbf{48.52} & 1 & N/A \\
\cmidrule{2-8}
&$\maxsc$ & \textbf{84.98} & \textbf{99.73} & \textbf{74.03} & \underline{49.1} & 1 & N/A \\
\cmidrule{2-8}
&$\minvc$ & 84.80 & \textbf{99.73} & 73.75 & 49.23 & 50 & \textbf{0.037} \\
\cmidrule{2-8}
&$\minvpc$ & \underline{84.91} & \textbf{99.73} & \underline{73.91} & 49.11 & 2 & \underline{1.15} \\
\bottomrule
\end{tabular}
\caption{Result on optimal solutions under different cardinality-optimisation criteria over the datasets}
\label{tbl:opt-c}
\end{table}
\titlep{Accuracy}
For the simpler datasets, \imdb, \corimdb, and \music, we observed that all optimization criteria produced a single optimal solution identical to $\maxes$, resulting in the same $\overline{F}_1$ (see Appendix \ref{app:simpler}). This is consistent with findings from~\cite{XiangBCGI24}, indicating that in the presence of data with less corrupted values and nulls, value-based comparisons (e.g., similarity or equality) are sufficient for merging, leaving little room for variation.
In contrast, for more complex datasets, $\minvps$ and $\minvpc$ achieved the highest accuracy on \cormusic and \cellmusic, outperforming the default $\maxes$ by 1.5\% and 1.28\%, respectively. However, on \poke, the opposite trend was observed: $\maxes$ slightly outperformed $\minvps$ and $\minvpc$ by 0.36\% and 0.07\%, resp.
\\
We also observed that solutions based on different optimization criteria differ in the following:
\\
\titleit{minA \& minV vs maxS \& maxE} 
Except for a tie on \poke, $\minv$ and $\minvp$ consistently outperformed $\maxs$ and $\maxe$ in terms of precision. The largest improvements were observed on \cormusic and \cellmusic, with leads of 2.93\% and 2.7\% for the S-criteria, and 1.5\% and 1.9\% for the C-optimal solutions, respectively. In contrast, $\maxs/\maxe$ obtained better recall, with the largest leads of 1.73\% seen when comparing $\maxes/\mathsf{SS}$ to $\minvs$ on \poke. This highlights the different characteristics between $\minv/\minvp$ and $\maxs/\maxe$, while the former prioritizes precision, the latter is more inclusive and has better coverage. As a result, on datasets with more corrupted values like \cormusic and \cellmusic, $\minv/\minvp$ obtained higher $\overline{F}_1$.
\\
\titleit{maxE vs maxS \& minA vs minV}
$\overline{F}_1$ scores for $\maxe$ and $\maxs$ are very close, with only marginal differences on \cormusic and \cellmusic, and identical results on \poke. Although $\maxs$ typically achieves higher precision, $\maxe$ consistently yields the best recall. This suggests that although both criteria are similar in nature, prioritizing rule applications may lead to more cautious merging decisions while maintaining good coverage.
Similar trends were observed between $\minv$ and $\minvp$, where $\minvp$ achieved higher precision and better $\overline{F}_1$ scores on all datasets except \cormusic.
\\
\titleit{Set vs Cardinality}
Accuracy metrics between S-optimal and C-optimal solutions differ only slightly, with the former generally achieving higher precision and the latter showing better recall. Although S optimization often produces many solutions, C optimization solutions may be preferable when a single representative model is desired.

\titlep{Efficiency}
For S-optimization,  first-model solving time $\tenumone$ is generally at the same level across criteria, with the largest difference being 1.1 times between $\minvps$ and $\maxes$ on \poke. Enumeration speed $\tenumn$ is also similar, except on \poke, where $\minvps$ was 6 times faster than $\minvs$.
In contrast, C-optimization methods show greater variability. $\minvc$ was consistently the slowest for first-model computation, taking up to 2.8, 1.76, and 1.01 times longer than the fastest criteria ($\maxec$ and $\maxsc$) on \cormusic, \cellmusic, and \poke, resp.
\\
Comparing the solving times of each S-optimal criterion with its C-optimal counterpart, we observed substantial increases in both $\tenumone$ and $\tenumn$—despite executing C-optimal computations on 36 threads. Without multi-threaded solving, both $\minvc$ and $\minvp$ hit the 24-hour timeout, even with heuristic-driven solving enabled (see Appendix \ref{app:exp}). These results suggest that in resource-constrained environments, S-optimal solutions are more practical, as the marginal quality gains from C-optimal solutions often come at a significant computational cost.



\subsection{Qualitative Studies}
We present qualitative examples from the
\cellmusic dataset, illustrating the efficacy and necessity of local semantics.
\\
In the \cellmusic dataset, the schema includes:
\begin{itemize}
    \item \textit{Place}(\textit{pid},\textit{name},\textit{type},\textit{address},\textit{area},\textit{coordinate}): stores information of building or outdoor area used for performing or producing music, where the \textit{area} attribute is a reference to \textit{id} of the \textit{Area} table. 
    \item $\textit{Release}(\textit{rid},\textit{artist},\textit{name},\textit{barcode},\textit{language})$: records information of release of music albums.
\end{itemize}
 The ER program for \cellmusic contains:
\enumi~a hard rule $hr_1$: the \textit{rid} of two releases are the same if they are from the same \textit{artist} and in the same \textit{language} and similar \textit{name}s.
\enumii~a soft rule $sr_1$: the \textit{pid} of two places are possibly the same if they have the same \textit{type} and similar \textit{name}s.
\enumiii~two denial constraints $d_1$ and $d_2$: FDs requiring that \textit{pid}/\textit{rid} must associate to the same \textit{coordinate}/\textit{barcode}, respectively.

\titlep{Block of merges} The two \textit{Place} tuples 
\begin{itemize}
    \item $\textit{Place}(t_1,p_1,\textit{Kunsthalle },\textit{Kindikty},a_1,\textit{48.20 81.63})$
    \item $\textit{Place}(t_2,p_2,\textit{Kunstha1le },\textit{Kindikty dist.},a_2,\textit{48.20-81.63})$ 
\end{itemize}
satisfy $sr_1$, but will be blocked by $d_1$ if the two \emph{value constants} \textit{48.20 81.63} and \textit{48.20-81.63} are not merged. Thus, the pair is absent in the stable models of \prev, but might be present in those of \ours.

\titlep{Violation of DCs} The two \textit{Release} tuples 
\begin{itemize}
    \item $\textit{Release}(t_3,r_1,at_1,\textit{chante les poètes},\textit{48.20 81.63},\textit{FR})$
    \item $\textit{Release}(t_4,r_2,at_2,\textit{chanteLes poètes},\textit{48.20-81.63},\textit{Fr})$ 
\end{itemize}
satisfy the hard rule $hr_1$, requiring that $\mathtt{eqo(r1, r2)}$ be included in any stable model. However, without merging the value constants \textit{48.20 81.63} and \textit{48.20-81.63}, the program violates the DC $d_2$, resulting in no solution for \cellmusic. As a result, $d_2$ must be excluded from \prev’s ER program. In contrast, \ours can safely incorporate $d_2$ by handling alternative representations of values.

\titlep{Local merges are necessary}
To satisfy $d_2$, the barcodes in the two \textit{Release} tuples must be merged. However, these same values also appear as \textit{coordinates} in the \textit{Place} table, where they refer to distinct locations—one in Germany, the other in Kazakhstan. As a result, the merge must be handled locally using \ours; a global merge would incorrectly unify distinct places and propagate errors to related tables such as \textit{Area}, due to referential dependencies.

\section{Conclusions}
We have introduced \ours, a logic-based system for collective ER that supports both global and local merges, along with a suite of seven optimality criteria for preferred solutions. Our formalization and complexity analysis of these criteria offer new insights into optimal ER solution selection. Through extensive experiments, we demonstrate the practical benefits of local semantics and flexible optimization, achieving superior accuracy on complex, real-world datasets. \ours thus marks a significant step toward principled ER in noisy and multi-relational environments.

\section*{Acknowledgements}
The authors were partially supported by the ANR AI Chair INTENDED (ANR-19-CHIA-0014) and by MUR under the PNRR project FAIR (PE0000013). 
The authors thank the Potassco community  for their support in resolving the issues encountered while using \clingo.
\bibliographystyle{kr}
\bibliography{kr-sample}

\clearpage
\appendix
\section{Intuitions of Optimality Criteria}~\label{app:alt-opt}
\titlep{$\maxes/\maxec$.}
Under these criteria, solutions that merge more objects or values are preferred, as they tend to reduce duplication in data records. Set containment offers an intuitive basis for comparing solutions obtained through incremental applications of merging rules. In contrast, the $\maxec$ criterion can be useful in scenarios where two solutions overlap only partially, as it favors the one that achieves a greater total number of merges.

\titlep{$\maxsc$.}
Under this cardinality-based criterion, solutions that include the same number of merges can still be distinguished by taking into account the supporting rules. We use the $\suppset(E,V)$, which annotates merges with the rules that support them, to give greater weight to merges supported by multiple rules. Intuitively, duplicate entities or values can often be identified through different reasoning paths, so merges justified by several rules are considered more likely to correspond to true duplicates. Note that while we can also define a set-based criterion, $\maxss$, which takes into account the rules supporting merges, we have proven in Proposition 1 that $\maxes$ and $\maxss$ actually coincide.

\titlep{$\minvs/\minvc$.}
Under these criteria, solutions are compared based on the triggered merges that fail to perform — captured by the set $\vioset(E, V)$. Intuitively, we prefer solutions that minimize such 'missed' merges, as these indicate pairs that were active according to some rule but were not included in the final merge set.

 \titlep{$\minvps / \minvpc$.}
These criteria select solutions that minimize the number of violated soft rules, formalized by $\viorset(E, V)$. Intuitively, we prefer solutions that respect more soft rules, as these represent likely but non-mandatory matches. Minimizing violations promotes solutions that adhere more closely to the intended matching heuristics, possibly improving the overall quality and consistency of the resolved data.

\def\triv{\mathsf{triv}}
\def\hornCNF{\mathsf{Horn3SAT}}

\section{Proofs of Section~\ref{sec:framework}}
\label{sec:optrecp-proofs}

To simplify the presentation of the proofs, we introduce some additional notation and preliminary results:

\begin{itemize}
    \item We denote by $\dcons(D)$ the set of constants occurring in a database $D$.
    \item Given an $n$-ary query $q(x_1, \ldots, x_n)$ and $n$-tuple of constants $\vec{c}=(c_1, \ldots, c_n)$, we denote by $q[\vec{c}]$ the Boolean query obtained by replacing each $x_i$ by $c_i$.
    \item We use \emph{$\qvars(q)$} (resp.\ $\qcons(q)$) for the set of variables (resp.\ constants) occurring in a query $q$. With a slight abuse of notation, we sometimes denote by $\ratom\in q$ the fact that a relational atom $\ratom$ occurs in $q$. 
    \item Given an ER specification $\E$ and a database $D$, for any $X \in \{\maxes,\maxec,\maxss,\maxsc,\minvs,\minvc,\\
    \minvps,\minvpc\}$, we denote by $\MaxSOL_X(D,\E)$ the set of $X$-optimal solutions for $(D,\E)$.
\end{itemize}

The following definition, taken from~\cite[Definition~2]{BienvenuCGI23}, specifies how queries in rule bodies and constraints are to be evaluated over extended databases.

\begin{definition}\label{queryeval}
A Boolean query $q$ (possibly containing similarity and inequality atoms) is 
\emph{satisfied in} 
$\extdat{D}{\merO}{\merV}$, denoted $\extdat{D}{\merO}{\merV} \models q$, if there exists a function $h: \qvars(q) \cup \qcons(q) \rightarrow 2^{\dcons(D)} \setminus \{\emptyset\}$ and 
functions $g_\ratom: \{0, \ldots, k\} \rightarrow 2^{\dcons(D)}$ for each $k$-ary relational atom $\ratom\in q$, such that: 
\begin{enumerate}
\item\label{it:nonEmpty} $h$ is determined by the $g_\pi$:  for every $a \in \qcons(q)$, $h(a)=\{a\}$, and for every $z \in \qvars(q)$, $h(z)$ is the intersection of all sets $g_\ratom(i)$ such that $z$ is the $i$th argument of $\ratom$;
\item\label{it:atoms} for every relational atom $\ratom = R(u_0, u_1, \ldots, u_k) \in q$, $R(g_\ratom(0), g_\ratom(1), \ldots, g_\ratom(k)) \in \extdat{D}{\merO}{\merV}$,
and for every $1 \leq i \leq k$, if $u_i \in \qcons(q)$, then $u_i \in g_\ratom(i)$;
\item\label{it:ineqs} for every inequality atom $z \neq z' \in q$: $h(z) \cap h(z') = \emptyset$;
\item\label{it:sim} for every similarity atom $u \approx u' \in q$: there exist $c \in h(u)$ and $c' \in h(u')$ such that $c \approx c'$.
\end{enumerate}
For non-Boolean queries, the set $q(\extdat{D}{\merO}{\merV}\!)$ of answers to $q(\vec{x})$ contains those tuples $\vec{c}$ such that $\extdat{D}{\merO}{\merV} \models q[\vec{c}]$. 
\end{definition}

\begin{lemma}\label{lem:Eval}
    Let $q$ be a Boolean query (possibly involving similarity and inequality atoms), $D$ be a database, $E$ be an equivalence relation on $\ObjD(D)$, and $V$ be an equivalence relation on $\StrPos(D)$. Checking whether $\extdat{D}{E}{V} \models q$ can be done in polynomial time in the size of $D$, $E$, and $V$.
\end{lemma}

\begin{proof}
    First, it is immediate to see that computing $\extdat{D}{E}{V}$, \ie the extended database induced by $D$, $E$, and $V$, can be done in polynomial time in the size of $D$, $E$, and $V$. With a slight abuse of notation, let $\dom{\extdat{D}{E}{V}}$ be the set of \emph{sets of constants} occurring in the extended database $\extdat{D}{E}{V}$. Assuming that $q$ is fixed, we now show that checking whether $\extdat{D}{E}{V} \models q$ can be done in polynomial time.
    
    Observe that, due to Point~\ref{it:atoms} of Definition~\ref{queryeval}, if there exists functions $h$ and $g_{\ratom}$ for each relational atom $\ratom \in q$ witnessing that $\extdat{D}{E}{V} \models q$, \ie satisfying Points~\ref{it:atoms},~\ref{it:nonEmpty},~\ref{it:ineqs}, and~\ref{it:sim}, then the image of each $g_{\ratom}$ must necessarily be a subset of $\dom{\extdat{D}{E}{V}}$. It follows that, when seeking for functions $g_{\ratom}$ for each relational atom $\ratom \in q$ witnessing that $\extdat{D}{E}{V} \models q$, we can actually restrict the codomain of each $g_{\ratom}$ to $\dom{\extdat{D}{E}{V}}$ rather than considering $2^{\dcons(D)}$.

    Now, for each $k$-ary relational atom $\ratom \in q$, let $G_{\ratom}$ be the set of all possible functions $g_{\ratom}$ from $\{0, \ldots, k\}$ to $\dom{\extdat{D}{E}{V}}$. Since $q$ is assumed to be fixed, and therefore $k$ is fixed, the total number of such functions is only polynomial in the size of $D$, $E$, and $V$. Let $\ratom_1, \ldots, \ratom_n$ be the relational atoms occurring in $q$. From what said above, we have that $\extdat{D}{E}{V} \models q$ if and only if there exists functions $(g_{\ratom_1},\ldots,g_{\ratom_n})$ in the Cartesian product $G_{\ratom_1} \times \ldots \times G_{\ratom_n}$ such that, once computed the function $h$ as specified by Point~\ref{it:nonEmpty} of Definition~\ref{queryeval}, requirements~\ref{it:atoms},~\ref{it:ineqs}, and~\ref{it:sim}~are all satisfied. It is not hard to verify that \myi computing the function $h$ from $(g_{\ratom_1},\ldots,g_{\ratom_n})$ as specified by Point~\ref{it:nonEmpty}~and \myii checking whether requirements~\ref{it:atoms},~\ref{it:ineqs}, and~\ref{it:sim} are satisfied by $h$ and $(g_{\ratom_1},\ldots,g_{\ratom_n})$ can be both done in polynomial time. Since $q$ is assumed to be fixed, and therefore the cardinality of the Cartesian product $G_{\ratom_1} \times \ldots \times G_{\ratom_n}$ is only polynomial in the size of $D$, $E$, and $V$, we can immediately derive a polynomial time algorithm for checking whether $\extdat{D}{E}{V} \models q$ which just consider all the possible functions $(g_{\ratom_1},\ldots,g_{\ratom_n})$ in the Cartesian product $G_{\ratom_1} \times \ldots \times G_{\ratom_n}$. If for some $(g_{\ratom_1},\ldots,g_{\ratom_n}) \in G_{\ratom_1} \times \ldots \times G_{\ratom_n}$ requirements~\ref{it:atoms},~\ref{it:ineqs}, and~\ref{it:sim}~are all satisfied after computing $h$ as specified by Point~\ref{it:nonEmpty}, then we return \true (because $\extdat{D}{E}{V} \models q$); otherwise, we return \false (because $\extdat{D}{E}{V} \not\models q$). 
\end{proof}

Since rule bodies do not involve any kind of negation, a pair $\alpha$ that is active in $\extdat{D}{\merO}{\merV}$ w.r.t.~some rule $r$ 
remains active in $\extdat{D}{\merO'}{\merV'}$ 
for any possible pair $\tup{\merO',\merV'}$ such that $\merO \cup \merV \subseteq \merO' \cup \merV'$. In particular, this means that if $\alpha$ is used to construct a (candidate) solution $\tup{\merO,\merV}$ for $(D,\E)$, then $\alpha$ remains active w.r.t.~$\tup{\merO',\merV'}$. In informal terms: later merges cannot invalidate the reasons for performing an earlier merge. What said above is an immediate consequence of the next lemma.

\begin{lemma}\label{lem:monotone}
    Let $q(\vec{x})$ be an $n$-ary query (possibly involving similarity atoms), $D$ be a database, $\vec{c}$ be an $n$-tuple of constants, $E$ and $E'$ be two equivalence relations on $\ObjD(D)$, and $V$ and $V'$ be two equivalence relations on $\StrPos(D)$. If $\vec{c} \in q(\extdat{D}{E}{V})$ and $E \cup V \subseteq E' \cup V'$, then $\vec{c} \in q(\extdat{D}{E'}{V'})$.
\end{lemma}

\begin{proof}
    Suppose that $E \cup V \subseteq E' \cup V'$ and $\vec{c} \in q(\extdat{D}{E}{V})$, \ie $\extdat{D}{E}{V} \models q[\vec{c}]$. Since $E \cup V \subseteq E' \cup V'$, we can immediately derive the following: for each (extended) fact of the form $R(\{t\},C_1,\ldots,C_k)$ occurring in the extended database $\extdat{D}{E}{V}$, we have a corresponding (extended) fact of the form $R(\{t\},C'_1,\ldots,C'_k)$ occurring in the extended database $\extdat{D}{E'}{V'}$ such that $C_i \subseteq C'_i$ holds for every $1 \leq i \leq k$. Thus, since $q$ does not contain inequality atoms and since $\extdat{D}{E}{V} \models q[\vec{c}]$ holds by assumption, due to the previous observation and the notion of evaluation of Boolean queries over extended databases (Definition~\ref{queryeval}), it is not hard to see that $\extdat{D}{E'}{V'} \models q[\vec{c}]$ as well, \ie $\vec{c} \in q(\extdat{D}{E'}{V'})$.
\end{proof}

\begin{lemma}\label{lem:NoMoreMerg}
    Let $\delta$ be a denial constraint over a schema that do not use inequality atoms, $D$ be a database, $E$ and $E'$ be two equivalence relations over $\ObjD(D)$, and $V$ and $V'$ be two equivalence relations over $\StrPos(D)$. If $\extdat{D}{E}{V} \not\models \delta$ and $E \cup V \subseteq E' \cup V'$, then $\extdat{D}{E'}{V'}\not\models \delta$. 
\end{lemma}

\begin{proof}
    Let $\delta=\forall \vec{y}. \varphi(\vec{y}) \rightarrow \bot$. By definition, we have that $\extdat{D}{E}{V} \not\models \delta$ iff $\extdat{D}{\merO}{\merV} \models \exists \vec{y}.\varphi(\vec{y})$. So, suppose that $E \cup V \subseteq E' \cup V'$ and $\extdat{D}{E}{V} \not\models \delta$, \ie $\extdat{D}{E}{V} \models \exists \vec{y}.\varphi(\vec{y})$. Since by assumption $\delta$ is a denial constraint without inequality atoms, and therefore $\exists \vec{y}.\varphi(\vec{y})$ is a CQ (without inequality atoms), and since $\extdat{D}{E}{V} \models \exists \vec{y}.\varphi(\vec{y})$ and $E \cup V \subseteq E' \cup V'$, by Lemma~\ref{lem:monotone} we derive that $\extdat{D}{E'}{V'} \models \exists \vec{y}.\varphi(\vec{y})$ as well, and therefore $\extdat{D}{E'}{V'}\not\models \delta$, as required.
\end{proof}

We are now ready to provide the proof of Proposition~\ref{prop:rel} and the proofs of Theorem~\ref{conp-opt} and Theorem~\ref{restr}. We point out that the lower bounds claimed in Theorem~\ref{conp-opt} and Theorem~\ref{restr} are provided using schemas whose relation symbols contain only object positions. This allows us to make three simplifying assumptions: \myi for such schemas, we can consider databases without occurrences of $\tidD$ constants, i.e.~databases are not $\tidD$-annotated; \myii we can write ER specifications over such schemas as pairs $\spec=\tup{\rulesO,\denC}$ rather than as triples $\spec=\tup{\rulesO,\emptyset,\denC}$; and \myiii we can write solutions to ER specifications $\spec$ over such schemas simply as equivalence relations $\merE$ over $\dcons(D)=\ObjD(D)$ rather than as pairs $\tup{\merE,\emptyset}$ (i.e.~the setting is exactly as in~\cite{lace_2022}).

\medskip

\noindent\textbf{Proposition~\ref{prop:rel}}. \emph{All pairs of the defined criteria yield different sets of optimal solutions, except for $\maxes$ and $\maxss$.}

\begin{proof}
    \textbf{Claim 1:} $\maxes$ and $\maxss$ yield the same set of optimal solutions, i.e.~we have that $\MaxSOL_{\maxes}(D,\E)=\MaxSOL_{\maxss}(D,\E)$ holds for any ER specification $\E$ and database $D$.

    Let $\E$ be an ER specification, $D$ be a database, $E$ be an equivalence relation over $\ObjD(D)$, and $V$ be an equivalence relation over $\StrPos(D)$. We now show that $\tup{E,V} \in \MaxSOL_{\maxes}(D,\E)$ if and only if $\tup{E,V} \in \MaxSOL_{\maxss}$, thus proving $\MaxSOL_{\maxes}(D,\E)=\MaxSOL_{\maxss}(D,\E)$.

    \emph{Only-if part} Suppose $\tup{E,V} \not\in \MaxSOL_{\maxss}$, i.e.~there is $\tup{E',V'}$ such that $\tup{E',V'} \in \SOL(D,\E)$ and $\suppset(E,V) \subsetneq \suppset(E',V')$. By combining Lemma~\ref{lem:monotone} and the notion of solution (Definition~\ref{def:SOL}), it is easy to see that any $\tup{E'',V''}$ such that $\tup{E'',V''} \in \SOL(D,\E)$ satisfies the following: if $p \in E'' \cup V''$, then $(p,r) \in \activepairs(D,E'',V'',\E)$ for some rule $r$ in $\E$. Thus, from this observation, the definition of $\suppset(\cdot,\cdot)$, and the fact that $\suppset(E,V) \subsetneq \suppset(E',V')$, we immediately derive that $\merO \cup \merV \subsetneq \merO' \cup \merV'$. Since $\tup{E',V'}$ is such that $\tup{E',V'} \in \SOL(D,\E)$, we can conclude that $\tup{E,V} \not\in \MaxSOL_{\maxes}$ as well.

    \emph{If part} Suppose $\tup{E,V} \not\in \MaxSOL_{\maxes}$, i.e.~there is $\tup{E',V'}$ such that $\tup{E',V'} \in \SOL(D,\E)$ and $\merO \cup \merV \subsetneq E'\cup V'$. By combining Lemma~\ref{lem:monotone} and the fact that $\merO \cup \merV \subsetneq E'\cup V'$, we immediately derive that $\activepairs(D,E,V,\E) \subsetneq \activepairs(D,E',V',\E)$, and therefore $\suppset(E,V) \subsetneq \suppset(E',V')$. Since $\tup{E',V'}$ is such that $\tup{E',V'} \in \SOL(D,\E)$, we can conclude that $\tup{E,V} \not\in \MaxSOL_{\maxss}$ as well.

    \textbf{Claim 2:} Let $X$ be any criteria based on set inclusion (i.e.~$X \in \{\maxes,\minvs,\minvps\}$), and let $Y$ be any criteria based on cardinality (i.e.~$Y \in \{\maxec,\maxss,\minvc,\minvpc\}$). Then, there exists a specification $\E$ and a database $D$ such that $\MaxSOL_X(D,\E) \neq \MaxSOL_Y(D,\E)$.

    Consider $\E=\tup{\hsrulesO,\emptyset,\Delta}$, where $\hsrulesO=\{\sigma,\sigma'\}$ and $\Delta=\{\delta\}$ are defined as follows: 
    \begin{itemize}
        \item $\sigma=R(x,y) \sarrow \linkO(x,y)$
        \item $\sigma'=R'(x,y) \sarrow \linkO(x,y)$
        \item $\delta=\forall y,z.\neg(R(y,y) \wedge R'(z,z))$
    \end{itemize}
    Let $D=\{R(a_1,a_2), R'(b_1,b_2), R'(c_1,c_2)\}$. Observe that, due to $\delta$, no $\tup{E,\emptyset} \in \SOL(D,\E)$ can be such that $(a_1,a_2) \in E$ and either $(b_1,b_2) \in E$ or $(c_1,c_2) \in E$. Now, consider $E_{a}=\{(a_1,a_2)\}$ and $E_{b,c}=\{(b_1,b_2),(c_1,c_2)\}$, and let $E'=\eqrel(E_a,\ObjD(D))$ and $E''=\eqrel(E_{b,c},\ObjD(D))$. It is straightforward to verify that $\MaxSOL_{\maxes}(D,\E)=\MaxSOL_{\minvs}(D,\E)=\MaxSOL_{\minvps}(D,\E)=\{\tup{E',\emptyset},\tup{E'',\emptyset}\}$ while $\MaxSOL_{\maxec}(D,\E)=\MaxSOL_{\maxsc}(D,\E)=\MaxSOL_{\minvc}(D,\E)=\MaxSOL_{\minvpc}(D,\E)=\{\tup{E'',\emptyset}\}$.

    \textbf{Claim 3:} There exists a specification $\E$ and a database $D$ such that $\MaxSOL_{\maxes}(D,\E) \neq \MaxSOL_{\minvs}(D,\E)=\MaxSOL_{\minvps}(D,\E)$.

    Consider $\E=\tup{\hsrulesO,\emptyset,\Delta}$, where $\hsrulesO=\{\sigma,\sigma'\}$ and $\Delta=\{\delta\}$ are defined as follows: 
    \begin{itemize}
        \item $\sigma=R(x,y) \sarrow \linkO(x,y)$
        \item $\sigma'=\exists z. R(z,z) \wedge R'(x,y) \sarrow \linkO(x,y)$
        \item $\delta=\forall y,z.\neg(R(y,y) \wedge R'(z,z))$
    \end{itemize}
    Let $D=\{R(a_1,a_2), R'(b_1,b_2)\}$. Observe that no $\tup{E,\emptyset} \in \SOL(D,\E)$ can be such that $(a_1,a_2) \in E$ and $(b_1,b_2) \in E$ as well as no $\tup{E,\emptyset} \in \SOL(D,\E)$ can be such that $(b_1,b_2) \in E$ but $(a_1,a_2) \not\in E$. Now, consider $E_{\emptyset}=\emptyset$ and $E_{a}=\{(a_1,a_2)\}$, and let $E'=\eqrel(E_{\emptyset},\ObjD(D))$ and $E''=\eqrel(E_{a},\ObjD(D))$. It is straightforward to verify that $\MaxSOL_{\maxes}(D,\E)=\{\tup{E'',\emptyset}\}$ while $\MaxSOL_{\minvs}(D,\E)=\MaxSOL_{\minvps}(D,\E)=\{\tup{E',\emptyset},\tup{E'',\emptyset}\}$.

    \textbf{Claim 4:} There exists a specification $\E$ and a database $D$ such that $\MaxSOL_{\minvs}(D,\E)\neq \MaxSOL_{\minvps}(D,\E)$.

    Consider $\E=\tup{\hsrulesO,\emptyset,\Delta}$, where $\hsrulesO=\{\sigma_a,\sigma_b, \sigma_c, \sigma_{c'}\}$ and $\Delta=\{\delta\}$ are defined as follows: 
    \begin{itemize}
        \item $\sigma_a=R_a(x,y) \sarrow \linkO(x,y)$
        \item $\sigma_b=R_b(x,y) \sarrow \linkO(x,y)$
        \item $\sigma_c=R_c(x,y) \sarrow \linkO(x,y)$
        \item $\sigma_{c'}=\exists z. R_b(z,z) \wedge R_c(x,y) \sarrow \linkO(x,y)$
        \item $\delta=\forall y,z.\neg(R_a(y,y) \wedge R_c(z,z))$
    \end{itemize}
    Let $D=\{R_a(a_1,a_2), R_b(b_1,b_2), R_c(c_1,c_2)\}$. Observe that, due to $\delta$, no $\tup{E,\emptyset} \in \SOL(D,\E)$ can be such that $(a_1,a_2) \in E$ and $(c_1,c_2) \in E$. Now, consider $E_{a}=\{(a_1,a_2)\}$ and $E_{a,b}=\{(a_1,a_2),(b_1,b_2)\}$, and let $E'=\eqrel(E_a,\ObjD(D))$ and $E''=\eqrel(E_{a,b},\ObjD(D))$. Observe that $\vioset(E',\emptyset)=\{(b_1,b_2),(c_1,c_2)\}$ while $\vioset(E'',\emptyset)=\{(c_1,c_2)\}$. Thus, since $E'' \in \SOL(D,\E)$, we derive that $E' \not \in \MaxSOL_{\minvs}(D,\E)$ because $\vioset(E'',\emptyset) \subsetneq \vioset(E',\emptyset)$. Notice, however, that $\viorset(E',\emptyset)=\{((b_1,b_2),\sigma_b),((c_1,c_2),\sigma_c)\}$ while $\viorset(E'',\emptyset)=\{((c_1,c_2),\sigma_c),((c_1,c_2),\sigma_{c'})\}$. So, $((c_1,c_2),\sigma_{c'}) \in \viorset(E'',\emptyset)$ but $((c_1,c_2),\sigma_{c'}) \not\in \viorset(E',\emptyset)$. Thus, it is easy to see that $E' \in \MaxSOL_{\minvps}(D,\E)$. We conclude that $\MaxSOL_{\minvs}(D,\E)\neq \MaxSOL_{\minvps}(D,\E)$.

    \textbf{Claim 5:} There exists a specification $\E$ and a database $D$ such that $\MaxSOL_{\maxec}(D,\E)=\MaxSOL_{\maxsc}(D,\E) \neq \MaxSOL_{\minvc}(D,\E)=\MaxSOL_{\minvpc}(D,\E)$.

    Recall $\E$, $D$, $E'$, and $E''$ as defined in the proof of Claim 3. It is straightforward to verify that $\MaxSOL_{\maxec}(D,\E)=\MaxSOL_{\maxsc}(D,\E)=\{\tup{E'',\emptyset}\}$ while $\MaxSOL_{\minvc}(D,\E)=\MaxSOL_{\minvpc}(D,\E)=\{\tup{E',\emptyset},\tup{E'',\emptyset}\}$. 

    \textbf{Claim 6:} There exists a specification $\E$ and a database $D$ such that $\MaxSOL_{\maxec}(D,\E) \neq \MaxSOL_{\maxsc}(D,\E)$.

    Consider $\E=\tup{\hsrulesO,\emptyset,\Delta}$, where $\hsrulesO=\{\sigma,\sigma',\sigma''\}$ and $\Delta=\{\delta\}$ are defined as follows: 
    \begin{itemize}
        \item $\sigma=R(x,y) \sarrow \linkO(x,y)$
        \item $\sigma'=R'(x,y) \sarrow \linkO(x,y)$
        \item $\sigma''=R''(x,y) \sarrow \linkO(x,y)$
        \item $\delta=\forall y,z.\neg(R(y,y) \wedge R'(z,z))$
    \end{itemize}
    Let $D=\{R(a_1,a_2), R'(b_1,b_2), R''(b_1,b_2)\}$. Now, consider $E_{a}=\{(a_1,a_2)\}$ and $E_{b}=\{(b_1,b_2)\}$, and let $E'=\eqrel(E_a,\ObjD(D))$ and $E''=\eqrel(E_{b},\ObjD(D))$. It is straightforward to verify that $\MaxSOL_{\maxec}(D,\E)=\{\tup{E',\emptyset},\tup{E'',\emptyset}\}$ while $\MaxSOL_{\maxsc}(D,\E)=\{\tup{E'',\emptyset}\}$. This is so because $\suppset{(E'',\emptyset)}=\{((b_1,b_2),\sigma'),((b_1,b_2),\sigma'')\}=2$ while $\suppset{(E',\emptyset)}=\{((a_1,a_2),\sigma)\}=1$.

    \textbf{Claim 7:} There exists a specification $\E$ and a database $D$ such that $\MaxSOL_{\minvc}(D,\E)\neq \MaxSOL_{\minvpc}(D,\E)$.

    Recall $\E$, $D$, $E'$, and $E''$ as defined in the proof of Claim 4. Consider, moreover, $E_{b,c}=\{(b_1,b_2),(c_1,c_2)\}$, and let $E=\eqrel(E_{b,c},\ObjD(D))$. Observe that $\vioset(E,\emptyset)=\{(a_1,a_2)\}$, and recall that $\vioset(E'',\emptyset)=\{(c_1,c_2)\}$. Thus, $|\vioset(E,\emptyset)|=|\vioset(E'',\emptyset)|=1$, and therefore it is easy to see that $\MaxSOL_{\minvc}(D,\E)=\{\tup{E,\emptyset},\tup{E'',\emptyset}\}$. However, note that $\viorset(E,\emptyset)=\{((a_1,a_2),\sigma_a)\}$ while $\viorset(E'',\emptyset)=\{((c_1,c_2),\sigma_c),((c_1,c_2),\sigma_{c'})\}$, and therefore $|\viorset(E,\emptyset)| < |\viorset(E'',\emptyset)|$, which implies that $\tup{E'',\emptyset} \not \in \MaxSOL_{\minvpc}(D,\E)$. We conclude that $\MaxSOL_{\minvc}(D,\E) \neq \MaxSOL_{\minvpc}(D,\E)$.
\end{proof} 

\noindent\textbf{Theorem~\ref{conp-opt}}. \emph{For all seven optimality criteria, recognition of optimal solutions is $\conp$-complete in data complexity.}

\begin{proof}
    \textbf{Upper bound.} First, given an ER specification $\E$, a database $D$, an equivalence relation $\merO$ over $\ObjD(D)$, and an equivalence relation $\merV$ over $\StrPos(D)$, computing $\extdat{D}{E}{V}$ as well as the set $\activepairs(D,E,V,\E)$ can be clearly done in polynomial time in the size of $D$, $E$, and $V$ (i.e.~assuming that $\E$ is fixed). Based on this observation, given an ER specification $\E$, a database $D$, a binary relation $E$ on $\ObjD(D)$, and a binary relation $V$ on $\StrPos(D)$, for any $X \in \{\maxes,\maxec,\maxsc,\minvs,\minvc,\minvps,\minvpc\}$, we now show how to check whether $\tup{E,V} \not \in \MaxSOL_X(D,\E)$ in $\np$ in the size of $D$, $E$, and $V$. 
    
    First, note that, by definition, $\tup{E,V} \not \in \MaxSOL_X(D,\E)$ if and only if either $\tup{E,V} \not \in \SOL(D,\E)$ or there exists $\tup{E'',V''}$ such that $\tup{E',V'} \in \SOL(D,\E)$ and $\tup{E',V'}$ is strictly better than $\tup{E,V}$ according to the optimality criteria $X$. So, we first guess a pair $\tup{E',V'}$, where $E'$ is a set of pairs of object constants from  $\ObjD(D)$ and $V'$ is a set of pairs of cells from $\StrPos(D)$. Then, we check whether \myi $\tup{E,V} \not \in \SOL(D,\E)$ or \myii $\tup{E',V'} \in \SOL(D,\E)$ and $\tup{E',V'}$ is strictly better than $\tup{E,V}$ according to the optimality criteria $X$. If either condition \myi or condition \myii is satisfied, then we return \true; otherwise, we return \false. Correctness of the above procedure for deciding $\tup{E,V} \not \in \MaxSOL_X(D,\E)$ is trivial.
    
    Due to~\cite[Theorem~2]{BienvenuCGI23}, note that checking whether $\tup{E,V} \in \SOL(D,\E)$ can be done in polynomial time in the size of $D$, $E$, and $V$ (and therefore, also checking whether $\tup{E',V'} \in \SOL(D,\E)$ can be done in polynomial time in the size of $D$, $E$, and $V$ because both the guessed $E'$ and $V'$ are polynomially related to $D$). Furthermore, since checking whether $\tup{E',V'}$ is strictly better than $\tup{E,V}$ according to the optimality criteria $X$ is clearly feasible in polynomial time in the size of $D$, $E$, and $V$ due to the fact that computing $\activepairs(D,E,V,\E)$ (resp.~$\activepairs(D,E',V',\E)$) can be done in polynomial time in the size of $D$, $E$, and $V$ (resp.~$D$, $E'$, and $V'$), we conclude that checking whether $\tup{E,V} \not \in \MaxSOL_X(D,\E)$ can be done in $\np$ in the size of $D$, $E$, and $V$.
    
    \textbf{Lower bound.} We now give a reduction from the complement of $\SAT$. We use the following fixed schema:
    \begin{align*}
        \S=\{&R_{\fff}/3,~R_{\fft}/3,~R_{\ftf}/3,~R_{\ftt}/3,~R_{\tff}/3,~R_{\tft}/3,\\
        &R_{\ttf}/3,~R_{\ttt}/3,~V/1,~F/1,~T/1,~B/1\}
    \end{align*}
    
    Given an input $\varphi=c_1 \wedge \ldots \wedge c_m$ to $\SAT$ over variables $ x_1,\ldots,x_n$, where $c_i= \ell_{i,1} \vee \ell_{i,2} \vee \ell_{i,3}$, 
    we construct a database $D^{\varphi}$ that contains: 
    \begin{itemize}
        \item the fact $V(x_i)$ for each $i \in [1,n]$,
        \item the facts $T(1)$, $F(0)$, $B(0)$, and $B(1)$;
        \item for each clause $c$ occurring in $\varphi$ of the form $(\overline{x_k} \vee \overline{x_z} \vee \overline{x_w})$ (resp.~$(\overline{x_k} \vee \overline{x_z} \vee x_w)$, $(\overline{x_k} \vee x_z \vee \overline{x_w})$, $(\overline{x_k} \vee x_z \vee x_w)$, $(x_k \vee \overline{x_z} \vee \overline{x_w})$, $(x_k \vee \overline{x_z} \vee x_w)$, $(x_k \vee x_z \vee \overline{x_w})$, $(x_k \vee x_z \vee x_w)$), 
        the fact $R_{\fff}(x_k,x_z,x_w)$ (resp.~$R_{\fft}(x_k,x_z,x_w)$, $R_{\ftf}(x_k,x_z,x_w)$, $R_{\ftt}(x_k,x_z,x_w)$, $R_{\tff}(x_k,x_z,x_w)$, $R_{\tft}(x_k,x_z,x_w)$, $R_{\ttf}(x_k,x_z,x_w)$, $R_{\ttt}(x_k,x_z,x_w)$).
    \end{itemize}
    We also let $\merO_{\triv}$ be the following equivalence relation over $\ObjD(D^{\varphi})$: $\merO_{\triv}=\eqrel(\emptyset,\ObjD(D^{\varphi}))$.

    We now define the fixed ER specification $\Sigma_{\SAT}=\tup{\hsrulesO,\Delta}$, 
where $\hsrulesO$ contains a
single soft rule for objects: 
\begin{itemize}
\item $\sigma =V(x) \wedge B(y) \sarrow \linkO(x,y)$
\end{itemize}
and $\Delta$ contains the following denial constraints (all the variables are implicitly universally quantified):
 \begin{itemize}
            \item $\delta_0= \neg(F(y) \wedge T(y))$
            \item $\delta_1=\neg(R_{\fff}(y_1,y_2,y_3) \wedge T(y_1) \wedge T(y_2) \wedge T(y_3))$
            \item $\delta_2=\neg(R_{\fft}(y_1,y_2,y_3) \wedge T(y_1) \wedge T(y_2) \wedge F(y_3))$
            \item $\delta_3=\neg(R_{\ftf}(y_1,y_2,y_3) \wedge T(y_1) \wedge F(y_2) \wedge T(y_3))$\item $\delta_4=\neg(R_{\ftt}(y_1,y_2,y_3) \wedge T(y_1) \wedge F(y_2) \wedge F(y_3))$\item $\delta_5=\neg(R_{\tff}(y_1,y_2,y_3) \wedge F(y_1) \wedge T(y_2) \wedge T(y_3))$\item $\delta_6=\neg(R_{\tft}(y_1,y_2,y_3) \wedge F(y_1) \wedge T(y_2) \wedge F(y_3))$\item $\delta_7=\neg(R_{\ttf}(y_1,y_2,y_3) \wedge F(y_1) \wedge F(y_2) \wedge T(y_3))$\item $\delta_8=\neg(R_{\ttt}(y_1,y_2,y_3) \wedge F(y_1) \wedge F(y_2) \wedge F(y_3))$
            \item $\delta_9 =\neg(V(v) \wedge B(v) \wedge V(x) \wedge T(y) \wedge F(z) \wedge$ \\ $\phantom{d}\qquad\quad  x \neq y \wedge x \neq z)$
\end{itemize}

It is straightforward to verify that both $D^{\varphi}$ and $E_{\triv}$ can be constructed in $\LOGSPACE$ from $\varphi$. To conclude the proof, we show that, for any $X \in \{\maxes,\maxec,\maxsc,\minvs,\minvc,\minvps,\minvpc\}$, we have that $E_{\triv} \in \MaxSOL_X(D^{\varphi},\E_{\SAT})$ if and only if $\varphi$ is unsatisfiable.

\emph{If part} Suppose that $E_{\triv} \not\in \MaxSOL_X(D^{\varphi},\E_{\SAT})$, i.e.~there exists $E'$ such that $E' \in \SOL(D^{\varphi},\E_{\SAT})$ and $E'$ is strictly better than $E_{\triv}$ according to the optimality criteria $X$. By construction of $\E_{\SAT}$ and $D^{\varphi}$, and in particular the fact that $\sigma$ is the only rule in $\hsrulesO$, we have that $E'$ must contain at least either a pair of the form $(x_i,0)$ or a pair of the form $(x_i,1)$, for some $x_i \in \dcons(D^{\varphi})$. Due to the presence of $\delta_9$ in $\E_{\SAT}$, from the assumption that $E' \in \SOL(D^{\varphi},\E_{\SAT})$, the presence of this pair in $E'$ actually implies that either $(x_j,0) \in E'$ or $(x_j,1) \in E'$, for every $j \in [n]$. Furthermore, due to the presence of $\delta_0$ in $\E_{\SAT}$, we cannot have both $(x_j,0) \in E$ and $(x_j,1) \in E$ for some $j \in [n]$. 

We can thus define a valuation $\mu$ by setting $\mu(x_j)=0$ if $(x_j,0) \in E'$ and $\mu(x_j)=1$ if $(x_j,1) \in E'$, for every $j \in [n]$. Since $E' \in \SOL(D^{\varphi},\E_{\SAT})$, this means that also the denial constraints $\delta_1,\ldots,\delta_8$ are satisfied in $D^{\varphi}_{E'}$. Thus, we can conclude that $\mu$ is a valuation to the variables of $\varphi$ witnessing the satisfiability of $\varphi$.

\emph{Only-if part} Suppose that $\varphi$ is satisfiable, and let $\mu \colon \{x_0, \ldots, x_n\} \rightarrow \{0,1\}$ be a satisfying valuation. Let $E_{\mu}=\{(x_i,\mu(x_i)) \mid 1 \leq i \leq n\}$, and let $E'=\eqrel(E_{\mu},\ObjD(D^{\varphi}))$. Obviously, $E'$ is strictly better than $E_{\triv}$ according to the optimality criteria $X$. We now conclude the proof by showing that $E' \in \SOL(D^{\varphi},\E_{\SAT})$, which implies that $E_{\triv} \not \in \MaxSOL_{X}(D^{\varphi},\E_{\SAT})$.

By construction, $(0,1) \not \in E'$, and therefore $D^{\varphi}_{E'} \models \delta_0$. 
Furthermore, since by assumption the valuation $\mu$ satisfies all the clauses of $\varphi$, there cannot be any violations of the denial constraints $\delta_1$ to $\delta_8$, i.e.~$D^{\varphi}_{E'} \models \delta_k$, for every $k=1,\ldots,8$. The last denial constraint $\delta_9$ is also satisfied in $D^{\varphi}_{E'}$ because $y$ and $z$ must map to $1$ and $0$ respectively, and for every $x_i$, either $x_i=1$ or $x_i=0$ holds in $D^{\varphi}_{E'}$.
\end{proof}

\noindent\textbf{Theorem~\ref{restr}}. \emph{In the restricted setting, recognition of optimal solutions becomes $\PTIME$-complete in data complexity for the optimality criteria $\maxes$, $\minvs$, and $\minvps$, while it remains $\conp$-complete in data complexity for the optimality criteria $\maxec$, $\maxsc$, $\minvc$, and $\minvpc$.}

\begin{proof}
    \textbf{Upper bound $\PTIME$ cases.} For $\maxes$ the claim follows from~\cite[Theorem~2]{BienvenuCGI23}. We now provide the proof for the remaining two cases, namely $\minvs$ and $\minvps$. We start with $\minvs$.

    Given an ER specification $\E=\tup{\hsrulesO,\hsrulesV,\Delta}$ such that the denial constraints in $\Delta$ do not make use of inequality atoms, a database $D$, a binary relation $E$ on $\ObjD(D)$, and a binary relation $V$ on $\StrPos(D)$, we first check whether $\tup{E,V} \in \SOL(D,\E)$. If this is not the case, then we return \false and we are done. This can be done in polynomial time in the size of $D$, $E$, and $V$ due to~\cite[Theorem~2]{BienvenuCGI23}. If $\tup{E,V} \in \SOL(D,\E)$, we continue as follows. We compute $\extdat{D}{E}{V}$, the set $\activepairs(D,E,V,\E)$, and the set $\vioset(E,V)$, which can be clearly done in polynomial time in the size of $D$, $E$, and $V$ (i.e.~assuming that $\E$ is fixed).

    For each $p \in \vioset(E,V)$, we do the following. If $p$ is a pair of object constants, then we start with $E'\coloneqq \eqrel(E \cup \{p\},\ObjD(D))$ and $V' \coloneqq V$; otherwise (\ie $p$ is a pairs of cells), we start with $E' \coloneqq E$ and $V' \coloneqq \eqrel(V \cup \{p\},\StrPos(D))$. In both cases, we repeat the following step until a fixpoint is reached: if there is some pair $(o,o')$ such that either $(o,o') \not\in \vioset(E,V)$ and $(o,o') \in \vioset(E',V')$ or $(o,o') \in q(\extdat{D}{E}{V})$ for some hard rule for objects $q(x,y) \harrow \linkO(x,y) \in \HRO$, then we set $E'\coloneqq\eqrel(E' \cup \{(o,o')\},\ObjD(D))$. Similarly, if there is some pair $(\tup{t,i},\tup{t',i'})$ such that either $(\tup{t,i},\tup{t',i'}) \not\in \vioset(E,V)$ and $(\tup{t,i},\tup{t',i'}) \in \vioset(E',V')$ or $(t,t') \in q(\extdat{D}{E}{V})$ for some hard rule for values $q(x_t,y_t) \harrow \linkV(\tup{x_t,i},\tup{y_t,i'}) \in \HRV$, then we set $V'\coloneqq\eqrel(V' \cup \{(\tup{t,i},\tup{t',i'})\},\StrPos(D))$. Once the fixpoint is reached, by construction we have that $\vioset(E',V') \subsetneq \vioset(E,V)$. Then, we just check whether the resulting $\tup{E',V'}$ is such that $\tup{E',V'} \in \SOL(D,\E)$. If this is the case for some $p \in \vioset(E,V)$, then we return \false; otherwise, we return \true. 

    Clearly, the overall procedure runs in polynomial time in the size of $D$, $E$, and $V$. Furthermore, Lemma~\ref{lem:monotone} and Lemma~\ref{lem:NoMoreMerg} guarantee its correctness. Notice, indeed, that the procedure considers pairs $\tup{E',V'}$ such that $\vioset(E',V') \subsetneq \vioset(E,V)$ and that are obtained by `minimally' extending $\tup{E,V}$ while ensuring that all the hard rules are satisfied. Consider now one $\tup{E',V'}$ obtained as detailed in the fixpoint step of the procedure. If it is the case that $\extdat{D}{E'}{V'} \not \models \Delta$, then Lemma~\ref{lem:NoMoreMerg} guarantees that it is not possible to resolve the constraint violation by further extending $\tup{E',V'}$. Similarly, it is not possible to resolve the constraint violation by deleting some pair (or a set of pairs) that was originally in $E \cup V$, as Lemma~\ref{lem:monotone} and the notion of candidate solution (Definition~\ref{def:SOL}) guarantee that the resulting $\tup{E',V'}$ after such deletion would be such that $\vioset(E',V') \not\subsetneq \vioset(E,V)$.

    Finally, the case for $\minvps$ is very similar, as it is sufficient to replace $p \in \vioset(E,V)$ with $\{p \mid (p,r) \in \viorset(E,V)\}$ in the steps of the above procedure.

    \textbf{Lower bound $\PTIME$ cases.} For $\maxes$ the claim follows from~\cite[Theorem~2]{BienvenuCGI23}. We now provide the proof for the remaining two cases, namely $\minvs$ and $\minvps$. We start with $\minvs$. The proof can be obtained by slightly extending the proof sketch of~\cite[Theorem~1]{lace_2022}, presented in the appendix of that paper. Specifically, we give a reduction from $\hornCNF$, a well-known $\PTIME$-complete problem~\cite{BoGG97}. We recall that an input to $\hornCNF$ is a pair $(F,x_w)$, where $F=\bigwedge_{i=1}^m \lambda_i$ is a formula over variables $X=\{x_1,\ldots,x_n\}$ and $x_w \in X$. Each $\lambda_i$ is either a single propositional variable $x_h \in X$ or a clause of the form $(\neg x_j \vee \neg x_k \vee x_h)$, where $x_j,x_k,x_h \in X$. The problem is to decide whether $F \models x_w$.
    
    We use the fixed schema $\S=\{R/4,~C/2,~W/2\}$. Given an input $\phi=(F,x_w)$ to $\hornCNF$ over variables $x_1,\ldots,x_n$, we construct a database $D^{\phi}$ that contains: 
    
    \begin{itemize}
        \item the fact $C(c_1,c_2)$;
        \item the fact $W(x_w,x'_w)$;
        \item for each $i \in [n]$, if $\lambda_i$ is of the form $\lambda_i=x_h$, then we include in $D^{\phi}$ the facts $R(l_i,t,t,x_h)$ and $R(l_i,t,t,x'_h)$; otherwise, i.e.~$\lambda_i$ of the form $(\neg x_j \vee \neg x_k \vee x_h)$, we include in $D^{\phi}$ the facts $R(l_i,x_j,x_k,x_h)$ and $R(l_i,x'_j,x'_k,x'_h)$.
    \end{itemize}
    We also let $\merO_{\triv}$ be the following equivalence relation over $\ObjD(D^{\phi})$: $\merO_{\triv}=\eqrel(\emptyset,\ObjD(D^{\phi}))$. It is straightforward to verify that both $D^{\phi}$ and $E_{\triv}$ can be constructed in $\LOGSPACE$ from $(F,x_w)$.

    We now define the fixed ER specification $\Sigma_{\hornCNF}=\tup{\hsrulesO,\Delta}$, where $\hsrulesO$ contains a single soft rule for objects $\sigma$ and a single hard rule for objects $\rho$:
    \begin{itemize}
        \item $\sigma = C(x,y) \sarrow \linkO(x,y)$
        \item $\rho = \exists z_l,z_1,z_2,z.R(z_l,z_1,z_2,x) \wedge R(z_l,z_1,z_2,y) \wedge \\ \phantom{d}\qquad\qquad\quad\quad\,\,\, C(z,z) \harrow \linkO(x,y)$
    \end{itemize}
    and $\Delta$ contains the following denial constraint $\delta$:
    \begin{itemize}
        \item $\delta=\forall y.\neg(W(y,y))$
    \end{itemize}

    The soft rule $\sigma$ allows to merge $c_1$ and $c_2$. If such a merge is performed, then, for each variable $x_h$ such that there is a $\lambda_i$ in $F$ of the form $\lambda_i=x_h$, the hard rule $\rho$ forces the merge of $x_h$ with its copy $x'_h$, due to the presence of $R(l_i,t,t,x_h)$ and $R(l_i,t,t,x'_h)$. Subsequently, assuming that $c_1$ and $c_2$ have been merged, by a trivial induction argument on the derivation length of the variables in $F$, it can be easily shown that $F \models x_h$ if and only if the merge of $x_h$ with its copy $x'_h$ is forced by the hard rule $\rho$. Finally, the denial constraint $\delta$ forbids the merge of $x_w$ with its copy $x'_w$.

    With the above intuition in mind, we now conclude the proof by showing that $\merO_{\triv} \in \MaxSOL_{\minvs}(D^{\phi},\E_{\hornCNF})$ if and only if $F \models x_w$.

    \emph{If part} Suppose that $\merO_{\triv} \not\in \MaxSOL_{\minvs}(D^{\phi},\E_{\hornCNF})$, i.e.~there exists $E'$ such that $E' \in \SOL(D^{\phi},\E_{\hornCNF})$ and $\vioset(E',\emptyset) \subsetneq \vioset(E_{\triv},\emptyset)$. By construction, this implies that $(c_1,c_2) \in E'$, as $\vioset(E_{\triv},\emptyset)=\{(c_1,c_2)\}$. Notice that, since $(c_1,c_2) \in E'$ and $E' \in \SOL(D^{\phi},\E_{\hornCNF})$, the following holds: for every variable $x_h$ for which $F \models x_h$, we have that the hard rule $\rho$ has forced the merge of $x_h$ with its copy $x'_h$, i.e.~for every variable $x_h \in X$, we have $(x_h,x'_h) \in E'$ if and only if $F \models x_h$. Now, from the fact that $D^{\phi}_{E'} \models \delta$ because $E' \in \SOL(D^{\phi},\E_{\hornCNF})$, we must necessarily have that $(x_w,x'_w) \not\in E'$. Thus, we can conclude that $F \not\models x_w$.

    \emph{Only-if part} Suppose that $F \not\models x_w$. Let $E_{F}=\{(c_1,c_2)\} \cup \{(x_h,x'_h) \mid F \models x_h\}$, and let $E'=\eqrel(E_F,\ObjD(D^{\phi}))$. Notice that, since by assumption $F \not\models x_w$, we have that $(x_w,x'_w) \not\in E'$, and therefore $D^{\phi}_{E'} \models \delta$. Furthermore, by construction we have that $D^{\phi}_{E'} \models \HRO=\{\rho\}$, as we included in $E'$ all the pairs $(x_h,x'_h)$ such that $F \models x_h$. It follows that $E' \in \SOL(D^{\phi},\E_{\hornCNF})$. Finally, observe that $\vioset(E',\emptyset)=\emptyset \subsetneq \vioset(E_{\triv},\emptyset)=\{(c_1,c_2)\}$. Since $E' \in \SOL(D^{\phi},\E_{\hornCNF})$ and $\vioset(E',\emptyset) \subsetneq \vioset(E_{\triv},\emptyset)$, we derive that $\merO_{\triv} \not\in \MaxSOL_{\minvs}(D^{\phi},\E_{\hornCNF})$.

    We observe that the above proof works for the optimality criteria $\minvpc$ as well (and actually, for all the seven optimality criteria). To see this, note that, for every input $(F,x_w)$ to $\hornCNF$, one can easily see that $\MaxSOL_{\minvs}(D^{\phi},\spec_{\hornCNF})=\MaxSOL_{\minvpc}(D^{\phi},\spec_{\hornCNF})$.

    \textbf{Upper bound $\conp$ cases.} The claimed upper bound for all the $\conp$ cases follows from Theorem~\ref{conp-opt}. 

    \textbf{Lower bound for $\minvc$ and $\minvpc$.} We give a reduction from the complement of $\SAT$. We use the fixed schema $\S'=\S \cup \{H/2\}$, where $\S$ is as in the proof of Theorem~\ref{conp-opt}. Given an input $\varphi$ to $\SAT$ over variables $x_1,\ldots,x_n$, we construct a database $\D^{\varphi}=D^{\varphi} \cup \{H(c_1,c_2)\}$, where $D^{\varphi}$ is as in the proof of Theorem~\ref{conp-opt} while $c_1$ and $c_2$ are two fresh constants. We also let $E^0=\{(x_i,0) \mid 1 \leq i \leq n\}$ and $E=\eqrel(E^0,\ObjD(\D^{\varphi}))$.

    We now define the fixed ER specification $\Sigma'_{\SAT}=\tup{\hsrulesO,\Delta}$, where $\hsrulesO$ contains two soft rules for objects: 
    \begin{itemize}
        \item $\sigma =V(x) \wedge B(y) \sarrow \linkO(x,y)$
        \item $\sigma' = H(x,y) \sarrow \linkO(x,y)$
    \end{itemize}
and $\Delta$ contains the following denial constraints (all the variables are implicitly universally quantified):
 \begin{itemize}
            \item $\delta_0= \neg(F(y) \wedge T(y))$
            \item $\delta_1=\neg(R_{\fff}(y_1,y_2,y_3) \wedge T(y_1) \wedge T(y_2) \wedge T(y_3) \wedge$ \\ $\phantom{d}\qquad\quad H(z,z))$
            \item $\delta_2=\neg(R_{\fft}(y_1,y_2,y_3) \wedge T(y_1) \wedge T(y_2) \wedge F(y_3) \wedge$ \\ $\phantom{d}\qquad\quad H(z,z))$
            \item $\delta_3=\neg(R_{\ftf}(y_1,y_2,y_3) \wedge T(y_1) \wedge F(y_2) \wedge T(y_3) \wedge$ \\ $\phantom{d}\qquad\quad H(z,z))
            $\item $\delta_4=\neg(R_{\ftt}(y_1,y_2,y_3) \wedge T(y_1) \wedge F(y_2) \wedge F(y_3) \wedge$ \\ $\phantom{d}\qquad\quad H(z,z))
            $\item $\delta_5=\neg(R_{\tff}(y_1,y_2,y_3) \wedge F(y_1) \wedge T(y_2) \wedge T(y_3) \wedge$ \\ $\phantom{d}\qquad\quad H(z,z))$
            \item $\delta_6=\neg(R_{\tft}(y_1,y_2,y_3) \wedge F(y_1) \wedge T(y_2) \wedge F(y_3) \wedge$ \\ $\phantom{d}\qquad\quad H(z,z))$
            \item $\delta_7=\neg(R_{\ttf}(y_1,y_2,y_3) \wedge F(y_1) \wedge F(y_2) \wedge T(y_3) \wedge$ \\ $\phantom{d}\qquad\quad H(z,z))
            $\item $\delta_8=\neg(R_{\ttt}(y_1,y_2,y_3) \wedge F(y_1) \wedge F(y_2) \wedge F(y_3) \wedge $ \\ $\phantom{d}\qquad\quad H(z,z))$
\end{itemize}

It is straightforward to verify that both $\D^{\varphi}$ and $E$ can be constructed in $\LOGSPACE$ from $\varphi$. We start with the optimality criteria $\minvc$, and show that $E \in \MaxSOL_{\minvc}(\D^{\varphi},\E'_{\SAT})$ if and only if $\varphi$ is unsatisfiable. First, note that $|\vioset(E,\emptyset)|=n+1$, i.e.~there are $n+1$ active pairs in $\D^{\varphi}_{\merE}$ w.r.t.~rules in $\rulesO$ that are not included in $E$, namely: $(c_1,c_2)$ and $(x_i,1)$, for each $i \in [1,n]$. Furthermore, since $(c_1,c_2) \not \in \merE$, we trivially derive that $\D^{\varphi}_{\merE} \models \denC$, and therefore $\merE \in \SOL(\D^{\varphi},\spec'_{\SAT})$.

\emph{If part} Suppose that $\merE \not\in \MaxSOL_{\minvc}(\D^{\varphi},\spec'_{\SAT})$, i.e.~there exists $E'$ such that $E' \in \SOL(\D^{\varphi},\E'_{\SAT})$ and $|\vioset(E',\emptyset)| < |\vioset(E,\emptyset)|$. Due to the simultaneous presence of $\sigma$ and $\den_0$, it is easy to see that any $E'' \in \SOL(\D^{\varphi},\E'_{\SAT})$ must be such that $|\vioset(E'',\emptyset)| \ge n$. In other words, since by assumption $E' \in \SOL(\D^{\varphi},\E'_{\SAT})$ and $|\vioset(E',\emptyset)| < |\vioset(E,\emptyset)|$, we have that $\merE'$ must contain the pair $(c_1,c_2)$ (which can be merged due to $\sigma'$) as well as either $(x_i,0)$ or $(x_i,1)$, for each $i \in [1,n]$ (but not both due to $\den_0$). 

We can thus define a valuation $\mu$ by setting $\mu(x_j)=0$ if $(x_j,0) \in E'$ and $\mu(x_j)=1$ if $(x_j,1) \in \merE'$, for every $j \in [n]$. Since $(c_1,c_2) \in \merE'$, from the assumption that $\merE' \in \SOL(\D^{\varphi},\spec'_{\SAT})$, and in particular the fact that the denial constraints $\delta_1,\ldots,\delta_8$ are satisfied in $\D^{\varphi}_{\merE'}$, we can conclude that $\mu$ is a valuation to the variables of $\varphi$ witnessing the satisfiability of $\varphi$.

\emph{Only-if part} Suppose that $\varphi$ is satisfiable, and let $\mu \colon \{x_0, \ldots, x_n\} \rightarrow \{0,1\}$ be a satisfying valuation. Let $E_{\mu}=\{(x_i,\mu(x_i)) \mid 1 \leq i \leq n\} \cup \{(c_1,c_2)\}$, and let $E'=\eqrel(E_{\mu},\ObjD(\D^{\varphi}))$. Obviously, $|\vioset(E',\emptyset)|=n < |\vioset(E,\emptyset)|=n+1$. We now conclude the proof by showing that $E' \in \SOL(\D^{\varphi},\E'_{\SAT})$, which implies that $E \not \in \MaxSOL_{\minvc}(\D^{\varphi},\E'_{\SAT})$.
    
By construction, $(0,1) \not \in E'$, and therefore $\D^{\varphi}_{E'} \models \delta_0$. Furthermore, since by assumption the valuation $\mu$ satisfies all the clauses of $\varphi$, there cannot be any violations of the denial constraints $\delta_1$ to $\delta_8$, i.e.~$\D^{\varphi}_{E'} \models \delta_k$, for every $k=1,\ldots,8$. Thus, we conclude that $E' \in \SOL(\D^{\varphi},\E'_{\SAT})$.

We observe that the above proof works for the optimality criteria $\minvpc$ as well. To see this, note that, for every input $\varphi$ to $\SAT$, one can easily see that $\MaxSOL_{\minvc}(\D^{\varphi},\spec'_{\SAT})=\MaxSOL_{\minvpc}(\D^{\varphi},\spec'_{\SAT})$.

   \textbf{Lower bound for $\maxec$ and $\maxsc$.} It only remains to address the cases of the optimality criteria $\maxec$ and $\maxsc$. We do so by extending the reduction previously illustrated for $\minvc$ and $\minvpc$. Specifically, recall the schema $\S'$, the ER specification $\E'_{\SAT}$, and the database $\D^{\varphi}$. We define the following fixed schema $\S''$, database $\dat^{\varphi}$, and fixed ER specification $\spec''_{\SAT}=\tup{\rulesO'',\denC''}$: 
    
    \begin{itemize}
        \item $\schema''=\schema' \cup \{V'/1,~P/2\}$;
        \item Given an input $\varphi$ to $\SAT$ over variables $x_1,\ldots,x_n$, we construct a database $\dat^{\varphi}=\D^{\varphi} \cup \{V'(x'_i) \mid 1 \leq i \leq n\} \cup \{P(x_i,x'_i) \mid 1 \leq i \leq n\}$;
        \item We also let $E^0=\{(x_i,0) \mid 1 \leq i \leq n\} \cup \{(x'_i,1) \mid 1 \leq i \leq n\}$ and $E=\eqrel(E^0,\ObjD(\dat^{\varphi}))$;
        \item $\hsrulesO''=\{\sigma,\sigma',\sigma''\}$ and $\denC''=\{\den_0,\den_1,\den_2,\den_3,\den_4,\den_5,\den_6,\den_7,\den_8,\den''_9\}$, where $\sigma$, $\sigma'$, $\den_0$, $\den_1$, $\den_2$, $\den_3$, $\den_4$, $\den_5$, $\den_6$, $\den_7$, and $\den_8$ are as illustrated above in the ER specification $\spec'_{\SAT}$, while $\sigma''=V'(x) \wedge Q(y) \sarrow \linkO(x,y)$ and $\den''_9=\forall y.\neg (P(y,y))$. 
    \end{itemize}

    Intuitively, $\sigma''$ allows the copy $x'$ of each variable $x \in \vec{x}$ to merge with $0$ or $1$ (but noth both due to $\delta_0$), while $\den''_{9}$ says that a variable and its copy cannot merge with the same truth value. This guarantees that in each $E \in \MaxSOL_{\maxec}(\dat^{\varphi},\E''_{\SAT})$ the equivalence class containing the constant $0$ has the same cardinality as the equivalence class containing the constant $1$.

    It is straightforward to verify that both $\dat^{\varphi}$ and $E$ can be constructed in $\LOGSPACE$ from $\varphi$. We start with the optimality criteria $\maxec$, and show that $E \in \MaxSOL_{\maxec}(\dat^{\varphi},\E''_{\SAT})$ if and only if $\varphi$ is unsatisfiable. First, note that there are overall four equivalence classes in $E$, namely: $[0]=\{0\} \cup \{x_i \mid 1 \leq i \leq n\}$, $[1]=\{1\} \cup \{x'_i \mid 1 \leq i \leq n\}$, $[c_1]=\{c_1\}$, and $[c_2]=\{c_2\}$, and therefore $|E|=2\cdot(n+1)^2+2$. Furthermore, since $(c_1,c_2) \not \in \merE$, we trivially derive that $\dat^{\varphi}_{\merE} \models \denC''$, and therefore $\merE \in \SOL(\dat^{\varphi},\spec''_{\SAT})$.

\emph{If part} Suppose that $\merE \not\in \MaxSOL_{\maxec}(\dat^{\varphi},\spec''_{\SAT})$, i.e.~there exists $E'$ such that $E' \in \SOL(\dat^{\varphi},\E''_{\SAT})$ and $|E| < |E'|$. By construction, due to the presence of $\den''_9$, it is easy to see that any $E''\in \SOL(\dat^{\varphi},\E''_{\SAT})$ must be such that the equivalence classes $[0]$ and $[1]$ in $E''$ contain at most $n+1$ constants. In other words, since by assumption $E' \in \SOL(\dat^{\varphi},\E''_{\SAT})$ and $|E| < |E'|$, we have that $\merE'$ must contain the pair $(c_1,c_2)$ (which can be merged due to $\sigma'$) as well as either $(x_i,0)$ and $(x'_i,1)$ or $(x_i,1)$ and $(x'_i,0)$, for each $i \in [1,n]$.

We can thus define a valuation $\mu$ by setting $\mu(x_j)=0$ if $(x_j,0) \in E'$ and $\mu(x_j)=1$ if $(x_j,1) \in \merE'$, for every $j \in [n]$. Since $(c_1,c_2) \in \merE'$, from the assumption that $\merE' \in \SOL(\dat^{\varphi},\spec''_{\SAT})$, and in particular the fact that the denial constraints $\delta_1,\ldots,\delta_8$ are satisfied in $\dat^{\varphi}_{\merE'}$, we can conclude that $\mu$ is a valuation to the variables of $\varphi$ witnessing the satisfiability of $\varphi$.

\emph{Only-if part} Suppose that $\varphi$ is satisfiable, and let $\mu \colon \{x_0, \ldots, x_n\} \rightarrow \{0,1\}$ be a satisfying valuation. Let $E_{\mu}=\{(x_i,\mu(x_i)) \mid 1 \leq i \leq n\} \cup \{(c_1,c_2)\} \cup \{(x'_i,\neg(\mu(x_i))) \mid 1 \leq i \leq n\}$, where $\neg(\mu(x_i))=0$ if $\mu(x_i)=1$ and $\neg(\mu(x_i))=1$ if $\mu(x_i)=0$, for each $i \in [n]$. Let, moreover $E'$ be $E'=\eqrel(E_{\mu},\ObjD(\dat^{\varphi}))$. Obviously, $|E| < |E'|$. We now conclude the proof by showing that $E' \in \SOL(\dat^{\varphi},\E''_{\SAT})$, which implies that $E \not \in \MaxSOL_{\maxec}(\dat^{\varphi},\E''_{\SAT})$.
    
By construction, $(0,1) \not \in E'$, and therefore $\dat^{\varphi}_{E'} \models \delta_0$. Furthermore, since by assumption the valuation $\mu$ satisfies all the clauses of $\varphi$, there cannot be any violations of the denial constraints $\delta_1$ to $\delta_8$, i.e.~$\dat^{\varphi}_{E'} \models \delta_k$, for every $k=1,\ldots,8$. Finally, for no $i \in [n]$ we have that both $(x_i,0)$ and $(x'_i,0)$ occur in $E'$ as well as for no $i \in [n]$ we have that both $(x_i,1)$ and $(x'_i,1)$ occur in $E'$. Thus, $\dat^{\varphi}_{E'} \models \delta''_9$ holds as well. We can therefore conclude that $E' \in \SOL(\dat^{\varphi},\E''_{\SAT})$.

We observe that the above proof works for the optimality criteria $\maxsc$ as well. To see this, note that, for every input $\varphi$ to $\SAT$, one can easily see that $\MaxSOL_{\maxec}(\dat^{\varphi},\spec''_{\SAT})=\MaxSOL_{\maxsc}(\dat^{\varphi},\spec''_{\SAT})$.
\end{proof}

\section{ASP Encoding and Implementation}
\subsection{Reified Format of ASP Encoding}~\label{app:reified}
In addition to the encoding format described in Section~\ref{sec:asp-enc}, database atoms (facts) can also be represented in a \emph{reified format}. For example, the fact
\[
\mathtt{author}(\mathtt{t}, \mathtt{x}, \mathtt{n}, \mathtt{d}, \mathtt{p})
\]
can be transformed into the set of reified facts
\[
\{\mathtt{tuple}(\mathtt{author}, \mathtt{t}, 1, \mathtt{x}), \dots, \mathtt{tuple}(\mathtt{author}, \mathtt{t}, 4, \mathtt{p})\}.
\]
While the reified format provides a more general and flexible representation, our preliminary experiments indicate that it leads to longer grounding times.

\subsection{Workflow Illustration}
\begin{figure*}[!htb] 
    \centering
    \includegraphics[width=0.6\textwidth]{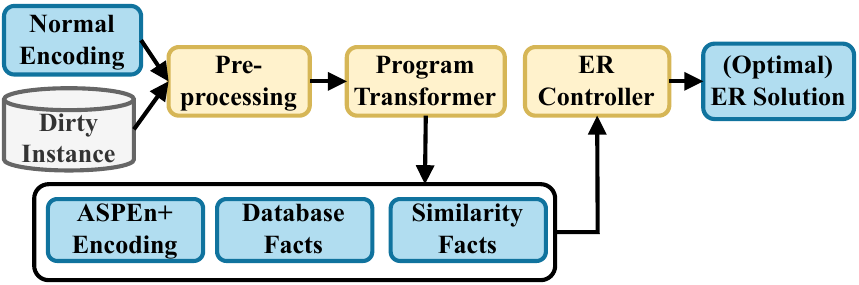}
    \caption{\ours's Workflow. Blue and yellow boxes indicate  ASP programs and \texttt{Python}-based procedures, respectively.
    }
    \label{fig:pip_new}
\end{figure*}
The workflow of \ours illustrated in~\fig{fig:pip_new}, 
begins with an ASP encoding of an ER specification, a \texttt{CSV}/\texttt{TSV} dataset containing duplicate and potentially corrupted values (e.g., nulls), and a set of command-line options. A preprocessing component computes similarity scores between relevant constant pairs based on attributes referenced in similarity atoms. 
These scores, along with the dataset, are passed to a program transformer, which rewrites the encoding into the \ours format and generates ASP facts for both database tuples and similarities. Finally, the ER controller, built on \texttt{clingo}, computes optimal solutions according to the specified input options.

\section{Experiments}~\label{app:exp}
\subsection{Experimental Setup}~\label{app:setup}
  
 \titlep{Datasets} Table~\ref{tbl:ds-stats} summarises the statistics of the datasets used in our experiments. Irrespective of the number of relations and attributes, datasets with more referential constraints exhibit greater structural complexity. Additionally, a higher number of distinct constants of datasets within the same domain (e.g. \music, \cormusic and \cellmusic) indicates a greater degree of syntactic variation of constants. The original data sources are available at:
\enumi~\imdb: \url{https://contribute.imdb.com/dataset},
\enumii~\music: \url{https://musicbrainz.org/doc/MusicBrainz\_Database/Schema},
\enumiii~\poke: \url{https://pokemondb.net/about}.
We note that all synthetic datasets were created from clean database instances, hence do not contain errors other than syntactic variants.

\begin{table}[!htp]
\renewcommand\arraystretch{0.8}
\setlength{\tabcolsep}{0.6em}
\centering
\begin{tabular*}{\linewidth}{@{}ccccccc@{}}
\toprule
\textbf{Name}& \textbf{\#Rec}& \textbf{\#Rel}& \textbf{\#At} &\textbf{\#Ref} &\textbf{\#Dup} &\textbf{\#Const}\\
\midrule
\imdb& 30k & 5& 22 & 4 & 6k & 64k\\
       \midrule
\corimdb& 30k & 5& 22 & 4 & 6k & 66k\\
       \midrule
\music& 41k & 11& 72 & 12 & 15k & 156k\\
    \midrule
\cormusic& 41k & 11& 72 & 12 & 15k & 160k \\
      \midrule
\cellmusic& 41k & 11& 72 & 12 & 15k & 166k \\
      \midrule
\poke & 240k & 20& 104 & 20 & 4k & 349k \\
\bottomrule
\end{tabular*}
\caption{Dataset Statistics.  \#-columns represent the number of records, relations, attributes, referential constraints and duplicates, respectively. 
}\label{tbl:ds-stats}
\end{table}

\titlep{Similarity Measures and Metrics}  We calculate the syntactic similarities of constants based on their data types~\cite{erblox-2017}.   (1) For numerical constants, we use the Levenshtein distance; (2) for short string constants (length$ <25$), we compute the score  as the editing distance of two character sequences (Jaro-Winkler distance);  (3) for long-textual constants (length$\geq 25$), we use the TF-IDF cosine score as the syntactic similarity measure.  We adopt  the commonly used metrics~\cite{DBLP-10} of \emph{Precision} (\textbf{P}), \emph{Recall} (\textbf{R}) and \emph{F1-Score} ($\mathbf{F_1}$) to examine the solutions derived from the specifications. Precision reflects the percentage of true merges in a solution and Recall indicates the coverage of true merges in a solution relative  to the ground truth. The quality of a solution is then measured as $\text{F}1 = 2\times\text{Precision}\times\text{Recall}/(\text{Precision}+\text{Recall})$.

\titlep{Environment}
We implemented \ours based on \texttt{clingo 5.5}~\footnote{https://potassco.org/clingo/python-api/5.5/} Python API. The ER programs follow the format of the  ASPCore2.0 standard~\cite{asp-core-2022}. All the experiments were run on a workstation using 3.8GHz  AMD Ryzen Threadripper 5965WX cores and 128 GB of RAM.

\subsection{Optimisation Configuration Benchmarking}~\label{app:bench}
We evaluated the runtime performance of different solver configurations across multiple optimality criteria and datasets.
\\
\titlep{Setup} For each dataset, we fixed the corresponding ER program and used \ours to compute between 1 and enumerate up to 50 optimal solutions, according to the criteria in Section~\ref{sec:opt-crt}, under a 128 GB RAM limit and a 5-hour time-out. For set-based optimization, we compared \textit{asprin} and domain-specific heuristics (\textit{heur}). For cardinality-based optimization, we evaluated: (i) optimization statements with multi-threaded solving using 36 CPU cores (\textit{wc-36}), (ii) optimization statements combined with domain-specific heuristics (\textit{wc+heur}), and (iii) \textit{asprin}.  We measured the solving time to compute the first optimal model ($t^1_s$) and the average solving time per model $t^n_s$, results are presented in Table~\ref{tbl:sopt} and Table~\ref{tbl:copt}.

\titlep{Results} For set-based optimization, \textit{heur} consistently outperformed \textit{asprin} across both $\tenumone$ and $\tenumn$ settings, often by substantial margins: it achieved speedups ranging from 1.3$\times$ to 15$\times$ on $\tenumone$ and from 146$\times$ to 909$\times$ on $\tenumn$. Notably, in both \cormusic and \cellmusic, \textit{asprin} failed to complete due to memory overflow (m/o), while \textit{heur} succeeded. These results indicate that \textit{heur} is a significantly more effective configuration for computing set-optimal solutions compared to \textit{asprin}.

A similar trend was observed for cardinality-based optimization, where \textit{asprin} was consistently outperformed by both \textit{wc+heur} and \textit{wc-36}, often by substantial margins. While \textit{wc+heur} frequently achieved better $\tenumone$ times than \textit{wc-36} (1$\times$ to 2$\times$ faster), and significantly outperformed it in $\tenumn$—notably by 629k$\times$ when computing $\maxsc$ on \cormusic—it failed to complete within the time limit (t/o) when solving for $\minvc$ or $\minvpc$ on more complex datasets, including \cormusic, \cellmusic, and \poke.
\begin{table*}[!htp]
        \renewcommand\arraystretch{0.5}
        \setlength{\tabcolsep}{0.7em}
        \centering
        \begin{tabular*}{\linewidth}{@{}ll|cc|cc|cc|cc|cc|cc@{}}
        \toprule
\multirow{2}{*}{\textbf{Opt}} & \multirow{2}{*}{\textbf{Met.}} 
& \multicolumn{2}{c}{\textbf{\imdb}} 
& \multicolumn{2}{c}{\textbf{\corimdb}} 
& \multicolumn{2}{c}{\textbf{\music}} 
& \multicolumn{2}{c}{\textbf{\cormusic}} 
& \multicolumn{2}{c}{\textbf{\cellmusic}} 
& \multicolumn{2}{c}{\textbf{\poke}} \\
& & $\tenumone$ & $\tenumn$ 
  & $\tenumone$ & $\tenumn$ 
  & $\tenumone$ & $\tenumn$ 
  & $\tenumone$ & $\tenumn$ 
  & $\tenumone$ & $\tenumn$ 
  & $\tenumone$ & $\tenumn$ \\
\midrule
  $\maxes/\mathsf{SS}$ & heur & \textbf{0.096 }& N/A &\textbf{ 0.69}  & N/A & \textbf{0.21} & N/A & \textbf{12.16}  & \textbf{1.86} & \textbf{23.02} &\textbf{ 1.67} & \textbf{48.52}  & N/A \\ 
 & asprin & 1.48 & N/A & 10.53 & N/A  & 3.16 & N/A & 174.24 & \rj{m/o} & 251.70 & \rj{m/o} & 76.51 & N/A \\ 
  \midrule
  $\minvs$  & heur & \textbf{0.18}  &N/A &\textbf{ 0.64}  & N/A & \textbf{0.44}  & N/A & \textbf{12.8} &\textbf{ 1.7 }& \textbf{23.01}  &\textbf{ 1.54} & \textbf{51.58}  & \textbf{0.06} \\ 
     & asprin& 1.28 & N/A & 10.21 & N/A& 3.20 & N/A & 170.26 & \rj{m/o}  & 237.99 &  \rj{m/o} & 76.49 & 8.78 \\ 
  \midrule
   $\minvps$  & heur & \textbf{0.71}   & N/A & \textbf{1.71}& N/A & \textbf{0.96} & N/A & \textbf{12.44} & \textbf{1.78} & \textbf{21.61}  & \textbf{2.48} & \textbf{56.87}  & \textbf{0.01} \\ 
& asprin & 3.77 & N/A &  13.62 & N/A & 4.73 & N/A & 175.04 & \rj{m/o}& 242.49 & \rj{m/o}& 74.78 & 9.09\\ 
        \bottomrule
        \end{tabular*}
        \caption{Comparisons of solving times for set-based optimisation across the datasets,  using different methods: heuristic (heur) and asprin.}
        \label{tbl:sopt}
        \end{table*}

\begin{table*}[!htp]
        \renewcommand\arraystretch{0.5}
        \setlength{\tabcolsep}{0.64em}
        \centering
        \begin{tabular*}{\linewidth}{@{}ll|cc|cc|cc|cc|cc|cc@{}}
        \toprule
\multirow{2}{*}{\textbf{Opt}} & \multirow{2}{*}{\textbf{Met.}} 
& \multicolumn{2}{c}{\textbf{\imdb}} 
& \multicolumn{2}{c}{\textbf{\corimdb}} 
& \multicolumn{2}{c}{\textbf{\music}} 
& \multicolumn{2}{c}{\textbf{\cormusic}} 
& \multicolumn{2}{c}{\textbf{\cellmusic}} 
& \multicolumn{2}{c}{\textbf{\poke}} \\
& & $\tenumone$ & $\tenumn$ 
  & $\tenumone$ & $\tenumn$ 
  & $\tenumone$ & $\tenumn$ 
  & $\tenumone$ & $\tenumn$ 
  & $\tenumone$ & $\tenumn$ 
  & $\tenumone$ & $\tenumn$ \\
\midrule
    & wc-36 & \underline{0.12} & N/A & \underline{0.79}  &  N/A & \underline{0.25}  &  N/A & \underline{35.0}9 & \underline{2.53} & \underline{66.71} & N/A & \textbf{48.52} & N/A\\ 
  $\maxec$ & wc+heur &\textbf{ 0.11} & N/A & \textbf{0.75} & N/A & \textbf{0.22 }& N/A & \textbf{23.05} & \textbf{1.75} & \textbf{33.77} & N/A & \underline{48.58}  & N/A \\ 
 & asprin & 1.57  & N/A & 10.55 & N/A  & 3.28 & N/A & 290.56 & \rj{m/o} & 287.27 & N/A& 74.90 & N/A \\ 
  \midrule
 & wc-36 &\underline{0.52}& N/A & \underline{1.24}  & N/A & \underline{0.97} & N/A & \textbf{30.1}  & \textbf{12.09} & \underline{52.8}  & \underline{629.56} & \underline{49.1}  & N/A \\ 
     $\maxsc$ & wc+heur &\textbf{ 0.41 }& N/A &\textbf{ 1.12} & N/A & \textbf{0.65} & N/A & \underline{57.80} & \underline{13.07} & \textbf{50.68} & \textbf{0.001} & \textbf{48.82} & N/A \\ 
     & asprin   & 6.68 &N/A & 17.49 &N/A & 9.64 & N/A & 264.55 & \rj{t/o} & 418.57 & \rj{t/o} & 60.35 & N/A \\ 
  \midrule
 & wc-36  & \textbf{0.18}  & N/A & \underline{0.89}  & N/A& \textbf{0.50} & N/A & \textbf{84.69} & \textbf{2.61}& \textbf{92.97} & \textbf{2.12}& \textbf{49.23} & \textbf{0.037}\\ 
  $\minvc$  & wc+heur & \underline{0.21} &N/A &\textbf{ 0.86} & N/A & \underline{0.61} & N/A & \rj{t/o} & \rj{t/o} & \rj{t/o}  & \rj{t/o} & \rj{t/o}  & \rj{t/o} \\ 
     & asprin& 1.81 & N/A & 11.22 & N/A& 3.60 & N/A & \rj{t/o} & \rj{t/o} & \rj{t/o} & \rj{t/o} & \rj{t/o} & \rj{t/o} \\ 
  \midrule
  & wc-36 &\underline{ 0.93 } & N/A &  \textbf{1.91}  & N/A & \textbf{1.22} & N/A &\textbf{48.97 }&\textbf{ 5.66} & \textbf{67.82} & \textbf{8.42} &\textbf{ 49.11}  & \textbf{1.15 }\\ 
   $\minvpc$  & wc+heur & \textbf{0.88} & N/A & \underline{2.01} & N/A & \textbf{1.22} & N/A & \rj{t/o} & \rj{t/o} & \rj{t/o} & \rj{t/o} & \rj{t/o}  & \rj{t/o}\\ 
& asprin &  4.24 & N/A &  14.52 & N/A & 5.52 & N/A & \rj{t/o}& \rj{t/o}& \rj{t/o} & \rj{t/o}& \rj{t/o} & \rj{t/o}\\ 
        \bottomrule
        \end{tabular*}
        \caption{Comparisons of solving times for cardinality-based optimisation across the datasets,  using different methods: 36-threaded weighted constraint (wc-36), combining heuristic and weighted constraint (wc+heur), and asprin.}
        \label{tbl:copt}
        \end{table*}

\subsection{Results on optimal solutions on simpler datasets}~\label{app:simpler}
For the simpler datasets, \imdb, \corimdb, and \music, we observe in Tables~\ref{tbl:opt-s-ex} and \ref{tbl:opt-c-ex} that all optimisation criteria produced a single optimal solution identical to $\maxes$, resulting in the same $\overline{F}_1$.
\begin{table}[htbp]
\renewcommand\arraystretch{0.15}
\setlength{\tabcolsep}{0.28em}
\centering
\begin{tabular}{cc|ccc|ccc}
\toprule
\textbf{Data} & \textbf{Method} & \multicolumn{3}{c|}{$\avgfone$~~~~~($\avgp$~~/~~~$\avgr$)} & $\tenumone$ & $\enumc$ & $\tenumn$ \\
\midrule
\multirow{4}{*}[-0.5ex]{\rotatebox{90}{\imdb}} 
& $\maxes/\mathsf{SS}$  &99.27 &99.39 & 99.14  & 0.096 & 1 & N/A \\
\cmidrule{2-8}
& $\minvs$  &99.27 &99.39 & 99.14   & 0.18 & 1 & N/A \\
\cmidrule{2-8}
& $\minvps$ &99.27 &99.39 & 99.14 & 0.71 & 1 & N/A \\
\midrule
\multirow{4}{*}[0.5ex]{\rotatebox{90}{\corimdb}} 
& $\maxes/\mathsf{SS}$ &  99.13 & 99.39 & 98.87 & 0.69 & 1 & N/A \\
\cmidrule{2-8}
& $\minvs$ &  99.13 & 99.39 & 98.87 & 0.64 & 1 & N/A \\
\cmidrule{2-8}
& $\minvps$ &99.13 & 99.39 & 98.87 & 1.71 & 1 & N/A \\
\midrule
\multirow{4}{*}[-0.5ex]{\rotatebox{90}{\music}} 
& $\maxes/\mathsf{SS}$ & 97.64 & 99.45 & 95.89 & 0.21 & 1 & N/A \\
\cmidrule{2-8}
& $\minvs$ & 97.64 & 99.45 & 95.89 &0.44& 1 & N/A \\
\cmidrule{2-8}
& $\minvps$ &97.64 & 99.45 & 95.89  & 0.96 & 1 & N/A \\
\bottomrule
\end{tabular}
\caption{Result on optimal solutions under different set-optimisation criteria over the datasets}
\label{tbl:opt-s-ex}
\end{table}

\begin{table}[htbp]
\renewcommand\arraystretch{0.15}
\setlength{\tabcolsep}{0.28em}
\centering
\begin{tabular}{cc|ccc|ccc}
\toprule
\textbf{Data} & \textbf{Method} & \multicolumn{3}{c|}{$\avgfone$~~~~~($\avgp$~~/~~~$\avgr$)} & $\tenumone$ & $\enumc$ & $\tenumn$ \\
\midrule
\multirow{4}{*}[-0.5ex]{\rotatebox{90}{\imdb}} 
& $\maxec$  &99.27 &99.39 & 99.14  & 0.12 & 1 & N/A \\
\cmidrule{2-8}
& $\maxsc$  &99.27 &99.39 & 99.14  & 0.52 & 1 & N/A \\
\cmidrule{2-8}
& $\minvc$  &99.27 &99.39 & 99.14   & 0.18& 1 & N/A \\
\cmidrule{2-8}
& $\minvpc$ &99.27 &99.39 & 99.14 & 0.93& 1 & N/A \\
\midrule
\multirow{4}{*}[0.5ex]{\rotatebox{90}{\corimdb}} 
& $\maxec$ &  99.13 & 99.39 & 98.87 & 0.79 & 1 & N/A \\
\cmidrule{2-8}
& $\maxsc$  &99.27 &99.39 & 99.14  & 1.24 & 1 & N/A \\
\cmidrule{2-8}
& $\minvc$ &  99.13 & 99.39 & 98.87 & 0.89 & 1 & N/A \\
\cmidrule{2-8}
& $\minvpc$ &99.13 & 99.39 & 98.87 & 1.91& 1 & N/A \\
\midrule
\multirow{4}{*}[-0.5ex]{\rotatebox{90}{\music}} 
& $\maxec$ & 97.64 & 99.45 & 95.89 & 0.25& 1 & N/A \\
\cmidrule{2-8}
& $\maxsc$ &97.64 & 99.45 & 95.89  & 0.97 & 1 & N/A \\
\cmidrule{2-8}
& $\minvc$ & 97.64 & 99.45 & 95.89 &0.50& 1 & N/A \\
\cmidrule{2-8}
& $\minvpc$ &97.64 & 99.45 & 95.89  & 1.22 & 1 & N/A \\
\bottomrule
\end{tabular}
\caption{Result on optimal solutions under different set-optimisation criteria over the datasets}
\label{tbl:opt-c-ex}
\end{table}
\clearpage
\end{document}